\documentclass[conference,compsoc]{IEEEtran}

\usepackage{silence}
\usepackage[T1]{fontenc}
\usepackage[noadjust,nocompress]{cite}
\usepackage[table]{xcolor}
\usepackage{tcolorbox}
\usepackage[highlightmode=immediate]{minted}
\usepackage{etoolbox}          %
\tcbuselibrary{minted}
\usepackage{listings}
\usepackage{lstlinebgrd}
\tcbuselibrary{listings}
\def\codethemewanted{github-light}
\def\codethemefallback{bw}
\setminted{fontsize=\small, breaklines, tabsize=4, autogobble}
\providecommand{\codemode}{pdf} 
\makeatletter
\newif\if@codetheme@ok
\long\def\codetheme@swallow#1{\global\@codetheme@okfalse}
\AtBeginDocument{%
  \@codetheme@oktrue
  \let\codetheme@savederr\minted@error
  \let\minted@error\codetheme@swallow
  \edef\codetheme@set{\noexpand\setminted{style=\codethemewanted}}\codetheme@set
  \setbox\z@=\hbox{\mintinline{text}{x}}%
  \let\minted@error\codetheme@savederr
  \if@codetheme@ok
  \gdef\codethemelexerprefix{github-}%
  \else
  \PackageWarning{minted}{Pygments style "\codethemewanted" not found; using
    monochrome fallback "\codethemefallback". Run 'uv sync' in ./minted (or
  enter the nix devShell) to enable the colour theme}%
  \edef\codetheme@set{\noexpand\setminted{style=\codethemefallback}}\codetheme@set
  \gdef\codethemelexerprefix{}%
  \fi
  \ifstrequal{\codemode}{pdf}{\usepdfcodetrue}{%
    \ifstrequal{\codemode}{minted}{\usepdfcodefalse}{%
  \if@codetheme@ok\usepdfcodefalse\else\usepdfcodetrue\fi}}%
}
\makeatother
\usepackage{booktabs}
\usepackage{multirow}
\usepackage{tabularx}
\usepackage{float}        %
\usepackage{amsmath}
\usepackage{newtxtext,newtxmath}
\usepackage[varqu,scaled=0.95]{inconsolata} %
\usepackage{pdfrender} %
\usepackage{algorithm}
\usepackage[noEnd=false, beginLComment={//\ }, endLComment={}]{algpseudocodex}
\usepackage{pgfplots}
\pgfplotsset{compat=1.18}
\usetikzlibrary{patterns, tikzmark, arrows.meta, calc}
\usepgfplotslibrary{groupplots}
\usepackage{hyperref}
\definecolor{darkblue}{HTML}{00008B}  %
\hypersetup{
  colorlinks=true,
  citecolor=darkblue,
  linkcolor=uchicago-maroon,
  urlcolor=black,
}
\usepackage{cleveref}
\crefname{appendix}{Appendix}{Appendices}
\Crefname{appendix}{Appendix}{Appendices}
\usepackage{caption}
\usepackage{subcaption}
\usepackage{xspace}
\usepackage{enumitem}
\usepackage{fontawesome5}      %
\usepackage{jie}               %

\newcommand{\langicon}[1]{\raisebox{-0.15em}{\includegraphics[height=1em]{assets/icons/#1}}}

\usepackage[
  activate={true,nocompatibility},
  final,
  protrusion=true,
  expansion=true,
  kerning=true,
  spacing=true,
  tracking=true,
  factor=1100,
  stretch=20,
  shrink=20
]{microtype}
\definecolor{gh-comment}{HTML}{6E7781}  %
\definecolor{diff-add}{rgb}{0.10, 0.55, 0.20}
\definecolor{diff-del}{rgb}{0.75, 0.15, 0.15}

\newcommand{\codeline}[1]{line~#1}

\newcommand{\agentlogo}[1]{\raisebox{-0.18ex}{\includegraphics[height=1em]{assets/figures/#1}}}
\newcommand{\claudelogo}{\agentlogo{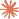}}
\newcommand{\codexlogo}{\agentlogo{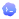}}

\definecolor{hlgreen}{HTML}{D5E8D4}       %
\definecolor{hlgreentext}{HTML}{000000}   %
\newtcbox{\ghl}{%
  on line,                 %
  nobeforeafter,
  boxsep=0pt, left=2pt, right=2pt, top=0.6pt, bottom=0.6pt,
  sharp corners,           %
  colback=hlgreen, colframe=black, boxrule=0.4pt,
  coltext=hlgreentext, fontupper=\bfseries,
}

\definecolor{hlred}{HTML}{F5B8B8}       %
\definecolor{hlredtext}{HTML}{000000}   %
\newtcbox{\rhl}{%
  on line,                 %
  nobeforeafter,
  boxsep=0pt, left=2pt, right=2pt, top=0.6pt, bottom=0.6pt,
  sharp corners,           %
  colback=hlred, colframe=black, boxrule=0.4pt,
  coltext=hlredtext, fontupper=\bfseries,
}

\NewDocumentCommand{\symbolon}{}{\mbox{{Symbolon}}\xspace}

\definecolor{deepgreen}{RGB}{0,150,0}
\newcommand{\gc}[1]{\cellcolor{deepgreen!#1}}

\NewDocumentCommand{\ie}{}{i.e.,\xspace}
\NewDocumentCommand{\eg}{}{e.g.,\xspace}

\newlist{rqlist}{enumerate}{1}
\setlist[rqlist]{label=\textbf{RQ\arabic*:}, ref=RQ\arabic*, leftmargin=3em}
\newif\ifinrqlist
\AddToHook{env/rqlist/begin}{\inrqlisttrue}
\AddToHook{env/rqlist/end}{\inrqlistfalse}
\NewDocumentCommand{\rqitem}{O{}}{%
  \ifinrqlist\else
  \PackageError{rqitem}%
  {\string\rqitem\space used outside rqlist environment}%
  {Wrap it in \string
    \begin{rqlist}...\string
  \end{rqlist}.}%
  \fi
\item\IfBlankTF{#1}{}{\textbf{#1}:\@}%
}

\makeatletter
\newcounter{note@pi}                                    %
\newcommand{\note@palette}[1]{\ifcase#1 %
  blue\or teal\or orange\or violet\or olive\or blue!55!red\or purple\or brown\or
red!70!black\or cyan!60!black\or magenta!70!black\or green!55!black\else black\fi}
\newcommand{\notecolor}[2]{\@namedef{note@c@#1}{#2}}    %
\newcommand{\note@ensure}[1]{%
  \@ifundefined{note@c@#1}{%
    \expandafter\xdef\csname note@c@#1\endcsname{\note@palette{\the\value{note@pi}}}%
    \stepcounter{note@pi}%
    \ifnum\value{note@pi}>11 \setcounter{note@pi}{0}\fi
}{}}
\def\note@ucfirst#1#2\@nil{\uppercase{#1}#2}            %
\newcommand{\note@render}[2]{%
  {\colorbox{\csname note@c@#1\endcsname}{\bfseries\sffamily\scriptsize\textcolor{white}{\note@ucfirst#1\@nil}}}%
  {\textcolor{\csname note@c@#1\endcsname}{\sf\small\(\blacktriangleright\)\textit{#2}~\(\blacktriangleleft\)}}%
}
\newcommand{\note@mkcmd}[1]{\csdef{#1}##1{\note@render{#1}{##1}}}
\newcommand{\note@register}[1]{%
  \edef\note@nm{\zap@space#1 \@empty}%
  \note@ensure{\note@nm}%
  \expandafter\note@mkcmd\expandafter{\note@nm}%
}
\newcommand{\reviewers}[1]{\forcsvlist{\note@register}{#1}}
\makeatother

\reviewers{penghui, kexin, ziyang, yizheng, jie, laurel}

\AtBeginDocument{}
\AtBeginDocument{}

\makeatletter
\RenewDocumentCommand{\paragraph}{}{\@startsection{paragraph}{4}{0pt}%
  {1ex \@plus .5ex \@minus .2ex}%
  {-0.5em}%
{\normalfont\normalsize\bfseries}}
\makeatother

\begin{document}

\title{\symbolon: Symbolic Execution by Learning Code Transformation}

\newcommand{\authorcell}[3]{%
  \begin{minipage}[t]{0.23\textwidth}\centering
    {\sublargesize #1}\\
    {\itshape #2\\#3}%
\end{minipage}}
\author{%
  \setlength{\tabcolsep}{3pt}%
  \begin{minipage}{\textwidth}\centering
    \begin{tabular}[t]{@{}cccc@{}}
      \authorcell{Jie Zhu}{University of Chicago}{jiezhu@uchicago.edu}     &
      \authorcell{Penghui Li}{Columbia University}{pl2689@columbia.edu}    &
      \authorcell{Zhongxuan Li}{University of Chicago}{zxli7@uchicago.edu} &
      \authorcell{Chihao Shen}{University of Maryland}{stevencs@umd.edu}
    \end{tabular}

    \vspace{13.4pt}

    \begin{tabular}[t]{@{}ccc@{}}
      \authorcell{Ziyang Li}{Johns Hopkins University}{ziyang@jhu.edu}  &
      \authorcell{Yizheng Chen}{University of Maryland}{yzchen@umd.edu} &
      \authorcell{Kexin Pei}{University of Chicago}{kpei@uchicago.edu}
    \end{tabular}
  \end{minipage}%
}

\maketitle
\thispagestyle{plain} %

\begin{abstract}
  Symbolic execution is a powerful program analysis technique with broad applications, such as vulnerability detection, security testing, and malware analysis.
However, this technique is known to suffer from scalability issues, \eg path explosion, complex constraints, due to certain structural and semantic patterns commonly presented in real-world programs.
Existing approaches attempt to escape these patterns by transforming programs into new representations to reduce the execution cost.
Unfortunately, these transformations are often too rigid to exploit diverse local program semantics and sometimes rely on compiler optimizations designed for concrete execution that may misalign with the goals of symbolic execution.

We present \symbolon, a framework that automatically learns diverse code transformations and applies them context-sensitively to improve symbolic execution.
Our key insight is to formulate transformation discovery as a search problem over program representations.
To make the search practical, \symbolon learns transformations cheaply offline on small programs, distills them into a reusable library of agent skills, and uses an agent to instantiate these skills on repo-level targets.
Our evaluation shows that \symbolon substantially improves the symbolic execution engine KLEE across 16 search strategies on 32 real-world programs, increasing line coverage by 3.69\(\times\) on average while reducing peak memory and per-query solver time by 29.2\(\times\) and 123\(\times\), respectively.
When applied to the latest Linux kernel, \symbolon uncovers 21 previously unknown bugs, all of which have been reported to the kernel maintainers.

\end{abstract}

\IEEEpeerreviewmaketitle

\section{Introduction}
\label{sec:introduction}

Symbolic execution is a well-established and influential program analysis technique that systematically explores program states under symbolic inputs\cite{king1976symbolic,cadar2011symbolic,cadar2008klee,chipounov2011s2e,poeplau2020symcc} to reason about program properties and behaviors\cite{coppa2017rethinking,schwerhoff2016advancing,hu2020automated,kuts2021towards}.
It has been used as a key building block for a wide range of code reasoning tasks, including checking reachability properties\cite{borzacchiello2025droidreach++,godefroid2005dart,person2011directed,chau2019analyzing,kuznetsov2012efficient,susag2022symbolic,baldoni2017assisting}, generating high-coverage test inputs\cite{cadar2008klee,stephens2016driller,chen2020savior,huang2020pangolin,hao2025syzspec,ling2025sound}, and detecting bugs and security vulnerabilities\cite{ruaro2021syml,shafiuzzaman2026guiding,ramos2015under,brown2020sys,wang2021maze,huang2022taming,zhang2024symbisect,shao2025cost,alharthi2025rakis,man2025scad,ding2024hunting}.
As a result, symbolic execution has supported production-scale testing workflows\cite{godefroid2012sage,bounimova2013billions,tillmann2008pex,tillmann2014transferring,godefroid2011higher}, and has been applied to analyze and secure a broad spectrum of real-world, security-critical software, including operating-system kernels and device stacks\cite{redini2017bootstomp,sun2022ksg,david2016binsec,ramos2015under,liu2022linkrid,han2023queryx,kim2022fuzzusb,hao2025syzspec}, browsers\cite{brown2020sys}, web and mobile applications\cite{li2014symjs,marques2025automated,loring2017expose,borzacchiello2022reach,liu2025domino,huang2025towards,artzi2008finding,artzi2010directed,santos2018symbolic,mirzaei2012testing}, commercial off-the-shelf (COTS) binary software\cite{poeplau2021symqemu,wang2017angr,cha2012unleashing,han2023queryx,qi2024symfit,daniel2020binsec}, low-level systems code\cite{chipounov2009selective,cebeci2024practical,iyer2024automatically}, embedded and firmware systems\cite{zaddach2014avatar,davidson2013fie,scharnowski2022fuzzware,liu2024co3}, and safety-critical cyber-physical systems\cite{kim2023patchverif,puasuareanu2010symbolic,yang2014directed,luckow2014exact,kurian2023automatically,kim2022reverse}.

Despite these successes, symbolic execution remains difficult to scale to large, real-world programs, where modest increases in program size and functionality can induce disproportionate cost due to path explosion and hard-to-solve path constraints\cite{cadar2013symbolic,baldoni2018survey,poeplau2020symcc,wei2023compiling,bounimova2013billions,bucur2011parallel}.
These costs often arise from common source-level program structures.
For example, when a loop bound depends on a symbolic input, the symbolic execution engine may fork many iterations, quickly exhausting its path budget\cite{cadar2013symbolic,cadar2008klee,baldoni2018survey,xiao2013characteristic,fromherz2017symbolic}.
Even along a single path, code that derives predicates or memory addresses from symbolic values, \eg bit-level encodings, pointer arithmetic, can produce expensive constraints for SMT solvers to discharge\cite{coppa2017rethinking,borzacchiello2019memory,xu2020benchmarking,chen2018learning}.
Importantly, these costs are \emph{representation-sensitive}.
Two equivalent implementations can expose different numbers of branches, different memory-access patterns, and different constraints for the solver, making them either easier or harder for symbolic execution to analyze\cite{zhang2024compiler}.
The scalability of symbolic execution is thus shaped not only by what a program computes, but also by how that \emph{computation is represented} in code.

This susceptibility to code representation suggests a complementary approach to improving symbolic execution by transforming and enriching the program representation before exposing it to symbolic execution\cite{poeplau2021symqemu,poeplau2020symcc,wei2020compiling,wei2023compiling}.
In fact, program transformation is known to substantially affect the behavior of program analyses\cite{van2020tailoring,namjoshi2018impact,jung2019fuzzification, peng2018tfuzz,guler2019antifuzz,zhu2025locus,zhang2023statfier}, and symbolic execution is no exception\cite{zhang2024compiler}.
Prior work has explored several ways to expose symbolic-execution-friendly code representations based on compiler optimizations\cite{dong2015studying,chen2018learning,zhang2024compiler}, targeted transformations for data structures and control flow\cite{cadar2015targeted,saumya2026taming,zhu2024loopscc}, operator-specific rewrites\cite{bouras2026defusing,perry2017accelerating,barr2018indexing}, and even semantics-breaking (behavior-changing) transformations that trade program equivalence for higher test coverage\cite{converse2017non}.

However, existing transformation-based approaches remain limited by how transformations are obtained and applied.
Compiler optimizations provide only a fixed set of transformations designed primarily for concrete execution objectives, such as execution time and code size, but they can misalign with symbolic execution objectives, such as optimizing coverage, reachability, and solver cost.
Other transformations tailored for symbolic execution are typically hand-coded for a narrow class of syntactic patterns or program constructs, \eg specific operators on symbolic indices, and thus require substantial human expertise, miss useful transformations outside predefined patterns, and are often too \emph{rigid} to be broadly applied.
This is problematic because the benefit of a transformation is highly sensitive to \emph{local context} (\Cref{sec:overview/motivating-example}).
These fixed transformations that simplify branches, insert predicates\cite{trabish2026enhancing,he2023eunomia,bouras2026defusing,crawford1996symmetry, gent1999symmetry}, or replace data structure operations, may remove a symbolic-execution barrier in one local context but suppress useful behavior in another, \eg bypassing (adding predicates like \texttt{klee\_assume}\cite{kleeintrinsics}) hash or decryption checks may help a parser reach deeper payload-processing logic, but can miss the core behavior in cryptographic libraries.
Existing approaches lack a mechanism for automatically discovering diverse, context-sensitive code transformations optimized for symbolic execution.

\paragraph{Our approach}
We present \symbolon, a framework that automatically discovers and applies diverse, context-sensitive code transformations for symbolic execution.
Our key idea is to formulate transformation discovery as a search problem over program representations.
Given a program that introduces symbolic execution barriers, \symbolon searches for an alternative source-level representation that helps a symbolic execution engine generate inputs with improved metrics, \eg coverage, on the original program.
With this search target, \symbolon expands the search space by essentially relaxing the transformations to be semantics-breaking\cite{converse2017non}, rather than restricting the search within compiler optimizations or hand-written templates that strictly preserve the semantics.
Specifically, \symbolon evaluates each transformation with a transform-and-replay reward.
It runs symbolic execution on the transformed program, replays the generated tests on the original program, and uses their improvement on the original code to guide the search.
In this formulation, the transformed program essentially acts as a test-generation scaffold, while the original program remains the optimization target.
This allows \symbolon to explore a much larger transformation space, while keeping the search reward aligned with symbolic execution objectives on the original program.

Making this formulation practical requires managing two costs: the open-ended transformation space and reward evaluation.
As semantics-breaking transformations further expand the search space, simple code mutation is unlikely to efficiently reach useful transformations.
Computing the reward exacerbates the cost by expensive symbolic execution (for test generation) and replay.
\symbolon addresses the first by using large language models (LLMs) as a search prior, drawing on their recent successes in guiding evolutionary search over programs and large systems\cite{romera2024mathematical,sen2026kiss,novikov2025alphaevolve,cheng2025barbarians}.
It addresses the second by decoupling transformation discovery from deployment.
Specifically, \symbolon learns the transformations only from small programs\cite{li2022codecontests}, where symbolic execution and replay are cheap, and distills effective transformations into reusable rules as persistent agent skills\cite{xu2026agent,anthropic2025agentskills}.
These learned skills capture recurring transformation patterns, grounding later agentic transformations for larger, real-world projects.
As this learning incurs only a one-time cost, its cost can be amortized across projects and symbolic execution campaigns.
\Cref{fig:overview} shows \symbolon's workflow.

Such a technical design in \symbolon directly addresses the rigidity of prior transformation-based approaches.
The learned skills are discovered automatically rather than manually specified, instantiated in local contexts rather than applied globally, and selected by replay-based feedback rather than by potentially misaligned objectives in compiler optimizations.
As transformations are expressed at the source level, the agent can leverage rich source-level hints, \eg symbol names, comments, to adapt each transformation rule in the learned agent skills to the target context.
Importantly, the source-level transformations make \symbolon orthogonal to engine-side improvements like search strategies\cite{chen2022symsan,yun2018qsym,yao2025empc,ruaro2021syml,he2021learch,yi2024cbc,sun2024cgs,li2013sgs,tu2026cottontail} and solver optimizations\cite{luo2026concollmic,li2025large,tu2026cottontail}.
As a result, \symbolon consistently improves the effectiveness of KLEE across all its search strategies, \eg by 3.69\(\times\) in coverage (see \Cref{sec:evaluation}), without the need to adapt to any specific symbolic execution and compiler configurations.

\paragraph{Results}
We evaluate \symbolon across 32 popular real-world software projects and 16 symbolic execution search strategies, including KLEE's built-in strategies and the state-of-the-art searchers\cite{yao2025empc, he2021learch, yi2024cbc, sun2024cgs,li2013sgs}, to measure how \symbolon complements existing approaches.
\symbolon consistently improves coverage across all search strategies, achieving an impressive 3.69\(\times\) more covered lines, averaged across all projects, while significantly reducing symbolic execution overhead, \ie by 29.2\(\times\) and 123\(\times\) in peak memory usage and solver time per query, respectively.
It also uncovers more sanitizer-reported security violations across all search strategies and identifies 21 new bugs when used to augment the symbolic-execution-based Linux kernel bug-finding tool\cite{hao2025syzspec, syzkaller}.
We release our learned transformation rules (in agent skills), the agentic framework, and the transformed projects here: \repo{cirrus-uchicago/Symbolon}.

\paragraph{Contributions}
We make the following contributions:
\begin{itemize}[leftmargin=1.2em]
  \item We identify source-level program representation as the key optimization target for symbolic execution and formulate transformation discovery as an automated search problem over program representations.

  \item We present \symbolon, a framework that automatically learns diverse, reusable transformation rules from small programs and transfers them as persistent agent skills to facilitate diverse transformations of real-world projects.

  \item We operationalize replay on the original program as a verifiable reward for transformation search, allowing \symbolon to explore broad transformation spaces, including semantics-breaking ones, while ensuring the objective of transformations aligns with that of symbolic execution.

  \item We evaluate \symbolon across a broad range of real-world projects and search strategies, demonstrating consistent coverage improvements, a significant reduction in symbolic execution overhead, and strong bug-finding capabilities in systems such as the Linux kernel.
\end{itemize}

\section{Overview}
\label{sec:overview}

This section first describes the background of symbolic execution.
We then present an example to motivate our work.

\subsection{Symbolic Execution Background}
\label{sec:overview/symbolic-execution}
Symbolic execution is a program analysis technique for systematically reasoning about program behaviors\cite{king1976symbolic,cadar2008klee,cadar2011symbolic,poeplau2020symcc}.
Instead of running a program with concrete inputs, a symbolic execution engine runs program \(P\) with symbolic input to represent a set of possible inputs.
During execution, the symbolic execution engine maintains a set of execution states, where each state consists of the program counter that tracks the current execution progress, and the path constraints that collect accumulated conditions along the explored path to characterize the inputs capable of reaching that state.

When symbolic execution reaches a branch whose condition depends on symbolic values, it forks into multiple successor states, each extended with the corresponding branch predicate, and continues exploring different execution paths.
As the number of states grows exponentially with the number of forks, the symbolic execution engine needs to select which pending state to advance.
Prior work has proposed various strategies to prioritize certain states\cite{li2013sgs,he2021learch,yi2024cbc,sun2024cgs,yao2025empc}.
A state is removed when it becomes infeasible or reaches a termination statement, \eg program exit or assertion failure.
The symbolic execution engine translates the corresponding path constraints to a concrete test input that, when supplied to \(P\), drives execution along the same explored path.

\paragraph{Intrinsics}
Symbolic execution engines are typically equipped with a set of intrinsic functions to support symbolic analysis.
For example, \texttt{sym\_create()} introduces a symbolic variable into the current execution state, which is helpful when applying symbolic execution to analyze a specific function without starting from the program entry point.
In \Cref{sec:overview/motivating-example}, we use its variant \texttt{sym\_choose()} to create a symbolic integer that constrains the variable into a finite set of choices.
Another widely used intrinsic is \texttt{sym\_assume()}, which adds a new constraint to the current path constraint, \eg \texttt{sym\_assume(x > 0)} adds \texttt{x > 0} to the current path constraint.
\symbolon uses these intrinsics as part of the code transformation to make symbolic execution more efficient.

\subsection{Motivating Example}
\label{sec:overview/motivating-example}

\begin{figure*}[!t]
  \centering
  \includegraphics[width=\textwidth]{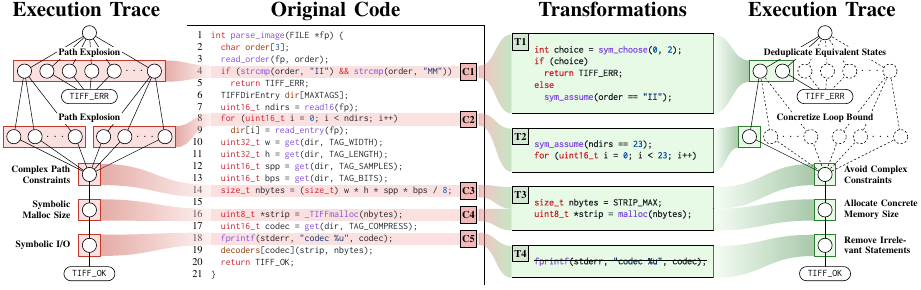}%
  \caption{A motivating example from \texttt{libtiff} showing how \symbolon helps symbolic execution by transforming the code. Each node in an execution trace represents an execution state, and each edge denotes a transition between two states. Nodes with dashed borders indicate states that are no longer explored in symbolic execution.}%
  \label{fig:motivating-examples}%
\end{figure*}

In \Cref{fig:motivating-examples}, we present a simplified code snippet from the \texttt{libtiff}\cite{libtiff} for parsing and checking the metadata of \texttt{TIFF} files.
A typical \texttt{TIFF} file starts with an 8-byte header that points to the first Image File Directory (IFD), where multiple IFDs can be linked together to support multi-page images.
Each IFD contains a 2-byte directory-entry count, followed by a sequence of 12-byte directory entries describing image metadata such as dimensions, sample layout, bit depth, and compression method, and ends with a 4-byte offset to the next IFD, or zero if there is no subsequent IFD.

\Cref{fig:motivating-examples} illustrates the key parsing logic that iterates over the directory entries in an input \texttt{TIFF} file and configures the image decoder based on the extracted compression method.
Specifically, the parser first validates the byte order (\codeline{4-5}), distinguishing between little-endian (\texttt{"II"}) and big-endian (\texttt{"MM"}). It then extracts the directory entry count \texttt{ndirs} for the current IFD (\codeline{7}) and loops through each entry \texttt{dir[i]} to retrieve its content (\codeline{8-9}).
After that, the parser extracts metadata fields and uses them to compute the image strip size \texttt{nbytes} (\codeline{14}), which determines the buffer size allocated to hold the image strip data.
Finally, it reads the compression tag and dispatches to the corresponding decoder (\codeline{16-20}).

\paragraph{Symbolic execution on the original code}
In this example, the harness supplies a fixed-size (\eg 300-byte) symbolic input file through \texttt{FILE *fp}.
The engine treats bytes read from \texttt{fp} as symbolic values and accumulates constraints over them.
After \texttt{read\_order} at \codeline{3}, the value of \texttt{order} is determined by the first two symbolic bytes, which encode the byte order of the \texttt{TIFF} file.
Branch \rhl{C1} checks whether this value matches one of the two valid byte-order markers.
Although this appears as a single source-level condition, the two \texttt{strcmp} calls compare strings byte by byte, causing symbolic execution to fork inside the library routine to distinguish different mismatch positions.
Most of these states return at \rhl{C1}.
Only states consistent with \texttt{order == "II"} or \texttt{order == "MM"} reach the subsequent parsing logic.

The next barrier is the directory-entry loop at \rhl{C2}.
The loop bound \texttt{ndirs} is read from the symbolic input at \codeline{7}, so the loop condition at \codeline{8} also depends on the symbolic data.
As a result, symbolic execution may fork at each iteration to represent whether the loop exits or continues, creating states for many possible counts of directory entries.
As \texttt{ndirs} is a 16-bit value, this loop can expose a large number of possible iterations and quickly exhaust the exploration budget before the engine reaches the following tag-processing logic.

Even after symbolic execution passes these path-explosion sources, the parser can still create solver-heavy expressions.
At \rhl{C3}, it computes \texttt{nbytes} by multiplying symbolic metadata fields such as width, height, samples per pixel, and bits per sample.
While the assignment itself does not add a path constraint, it builds a symbolic bit-vector expression.
When later operations constrain this value, the solver must reason about symbolic multiplication and C integer overflow semantics, which is often much harder than linear arithmetic over symbolic values\cite{cadar2013symbolic,de2008z3}.
Existing search strategies may reach this code earlier, but they do not simplify the expression used in these later queries.

The symbolic expression for \texttt{nbytes} then flows into another barrier at \rhl{C4}, where it is passed as the size argument to \texttt{\_TIFFmalloc}.
Many symbolic execution engines handle symbolic size allocations only under restrictions, \eg by concretizing the size, rejecting the allocation, or terminating states when the size cannot be bounded or modeled.
Therefore, exploration may stop before reaching the decoder dispatch.
Similarly, \rhl{C5} sends symbolic metadata to \texttt{fprintf}, forcing the engine to model formatted file I/O for an output side effect that is irrelevant to decoder selection.
These statements illustrate how basic memory management and I/O APIs can become barriers when they take symbolic values.

\paragraph{Symbolon transforms the original code}
The green snippets in \Cref{fig:motivating-examples} show how \symbolon transforms the original code into a representation that is easier for symbolic execution to analyze.
Note that these rewrites are not conventional compiler optimizations.
They rely on understanding the local source-level intent, such as which code implements a semantic check, which values are needed only to reach downstream parsing logic, and which side effects are irrelevant to the exploration goal.
Such intent is difficult for fixed compiler passes to exploit safely, and blindly optimizing for concrete execution can even obscure the symbolic execution objective.

Transformation \ghl{T1} replaces the byte-order check with an explicit symbolic choice that represents the semantic outcome of the check, \ie whether the byte order is valid or invalid.
This avoids forcing symbolic execution to reason through the byte-by-byte implementation in \texttt{strcmp} and instead exposes the source-level decision directly to the engine.
In the valid branch, \symbolon constrains \texttt{order} to \texttt{"II"} as a representative valid byte order, avoiding a separate exploration of the symmetric \texttt{"MM"} case.
This transformation is specific to the local context, in that it is appropriate here because the subsequent logic in the simplified example only needs a valid byte order to reach the tag-processing and decoder-dispatch code.
If later code were data-dependent on the distinction between \texttt{"II"} and \texttt{"MM"}, the transformation would need to preserve both valid cases.
This is similar in spirit to symmetry-breaking constraints\cite{crawford1996symmetry,gent1999symmetry} and streamlining\cite{gomes2004streamlined}, where additional constraints are introduced to avoid redundant regions of the search space.

Transformation \ghl{T2} addresses the symbolic loop bound at \rhl{C2}.
Instead of allowing the 16-bit symbolic value \texttt{ndirs} to induce many loop iterations, \symbolon rewrites the loop into a fixed-iteration one derived from the configured symbolic input size (in our harness, the symbolic \texttt{TIFF} input is capped at 300 bytes).
As a \texttt{TIFF} file uses an 8-byte header before the first IFD, and the IFD then contains a 2-byte entry count, a sequence of 12-byte directory entries, and a 4-byte pointer to the next IFD, the maximum number of directory entries that can fit in this input layout is thus \(\left\lfloor \frac{300 - 8 - 2 - 4}{12} \right\rfloor = 23\).
Since \texttt{tiffinfo} inspects metadata and does not need a large image payload to reach the downstream decoder-dispatch logic, this bound gives symbolic execution a concrete loop structure while preserving the ability to exercise the relevant parsing code on the original program during replay.

Transformations \ghl{T3} and \ghl{T4} remove downstream barriers that are not essential for reaching the decoder dispatch.
At \rhl{C3}--\rhl{C4}, the original code computes \texttt{nbytes} from symbolic metadata fields and passes it to \texttt{\_TIFFmalloc}, producing both an expensive symbolic expression and a symbolic size allocation.
\symbolon replaces \texttt{nbytes} in the transformed scaffold with the project-level bound \texttt{STRIP\_MAX}, giving the engine a concrete allocation size and avoiding the symbolic multiplication expression in later queries.
At \rhl{C5}, \symbolon removes the diagnostic \texttt{fprintf} call, which only prints the selected compression method and does not affect the parser state or the following decoder dispatch.
A standard compiler optimization generally cannot perform these rewrites, as concretizing \texttt{nbytes} or removing formatted output can change the program's observable behavior.
In \symbolon, however, since the transformed code is used only as a test-generation scaffold, all generated tests are replayed on the original program to measure improvement, with the original \texttt{nbytes} computation and \texttt{fprintf} call remaining intact.

Together, these transformations increase line coverage on \texttt{libtiff} from 1,251 to 1,660 lines (\(+32.7\%\)) during an one-hour symbolic execution campaign.
This comparison follows the same evaluation setting described in \Cref{sec:evaluation/setup}, where the line coverage is measured by replaying symbolic execution generated test cases on the original program.

\begin{figure*}[!t]
  \centering
  \includegraphics[width=\textwidth]{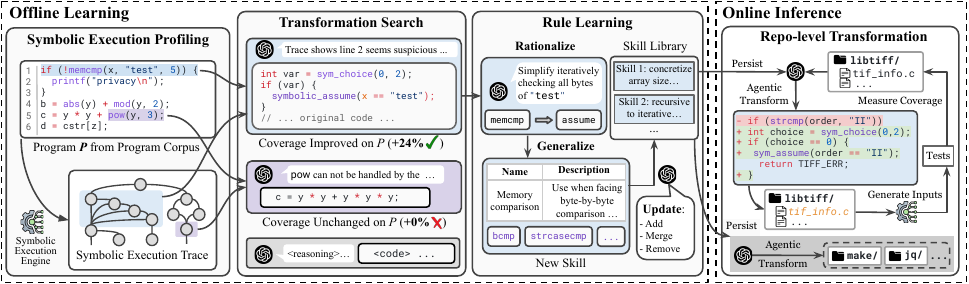}
  \caption{Overview of the \symbolon workflow. In the offline learning phase, \symbolon runs symbolic execution on simple programs and uses evolutionary search to learn a diverse collection of transformations. These transformations are then distilled into agent skills, which are further packaged into a reusable skill library for online inference to real-world programs. }
  \label{fig:overview}
\end{figure*}

\section{Methodology}
\label{sec:methodology}

\Cref{fig:overview} shows the workflow of \symbolon.
The goal of \symbolon is to learn source-level program representations that are more amenable to symbolic execution, while measuring progress on the original program.
This creates two practical challenges.
First, the space of possible representations is large, \eg useful transformations may summarize library checks, bound symbolic loops, rewrite arithmetic, insert assumptions, or remove exploration-irrelevant side effects.
Second, evaluating a transformation is expensive because it requires running symbolic execution on the transformed program and replaying the generated tests on the original program.
\symbolon addresses these challenges with a two-phase design.
In the offline learning phase, it searches for useful transformations on small programs, where symbolic execution and replay are cheap, and distills successful transformations into a skill library \(\mathcal{T}\).
In the repo-level transformation phase (online inference), it uses \(\mathcal{T}\) to adapt these learned transformations to a target project \(P_{\text{tgt}}\), producing a selected transformed scaffold \(P_{\text{sel}}\).
Symbolic execution is then run on \(P_{\text{sel}}\), and the generated tests are replayed on \(P_{\text{tgt}}\) to measure progress on the original code.

\subsection{Offline Learning}
\label{sec:methodology/transform-rule-learning}

The offline phase learns reusable transformation knowledge from small programs.
Its input is a corpus of programs \(\mathcal{P}\), and its output is a skill library \(\mathcal{T}\)\cite{anthropic2025agentskills}.
Each skill records a recurring symbolic-execution barrier, the source context in which it appears, and a transformation strategy that has been validated by replay on the original program.
The phase has three steps: profiling, transformation search, and skill distillation (rule learning).

\paragraph{Symbolic execution profiling}
Given a program \(P \in \mathcal{P}\), \symbolon first runs symbolic execution on \(P\) and collects a structured execution trace.
The trace records source locations reached by each state, branches that fork states, path constraints, solver queries, state-termination reasons, and coverage.
These events are mapped back to source locations, allowing \symbolon to identify code regions whose representation appears to block exploration.
For example, a byte-by-byte string comparison may create many short-lived states, and a symbolic multiplication may later produce expensive solver queries (\Cref{sec:overview/motivating-example}).

\symbolon then asks an LLM agent to propose candidate \emph{barriers} from the program and trace.
A barrier is a source-level explanation of why a code region is difficult for symbolic execution, not yet a trusted fact.
Candidate barriers are admitted into the skill library only if the later transformation search finds a rewrite whose generated tests improve replayed coverage on the original program.
Importantly, the LLM here only proposes hypotheses, while symbolic execution and replay provide the evidence.

\paragraph{Transformation search}
For each candidate barrier \(b\), \symbolon searches for an alternative source-level representation \(\widehat{P}\) of \(P\).
A transformed program \(\widehat{P}\) is useful if symbolic execution on \(\widehat{P}\) generates tests that improve a target metric when replayed on the original program \(P\).
In this paper, the main metric is coverage, but the same formulation can optimize other replay-measurable objectives such as reachability or bug-triggering behavior\cite{zhu2025locus}.

This search space can often be too broad for simple syntactic rewriting.
The same barrier may require different transformations depending on local context-sensitive features, \eg types, constants, library calls, data dependencies, and program intent.
As shown in \Cref{fig:motivating-examples}, for example, replacing a byte-order check with a representative valid case is only appropriate when the following code does not depend on the distinction between valid byte orders.
Similarly, concretizing a loop bound requires understanding the input layout.
Removing a diagnostic output is useful only when the output side effect is irrelevant to the exploration target.
\symbolon thus uses an LLM-guided evolutionary search\cite{openevolve,novikov2025alphaevolve,romera2024mathematical} to propose candidate representations.

Each transformed program \(\widehat{P}\) is evaluated by transform-and-replay.
\symbolon runs symbolic execution on \(\widehat{P}\), replays the generated tests on the original program \(P\), and computes the replayed metric on \(P\).
If the replayed metric improves over symbolic execution on \(P\), \symbolon keeps \((P,\widehat{P})\) as a validated transformation example for barrier \(b\).
As the reward is measured on \(P\), not on \(\widehat{P}\), the search can include semantics-breaking transformations while avoiding rewards for transformations that only make \(\widehat{P}\) appear to be easier.

\begin{algorithm}[!t]
  \caption{Offline learning of transformation skills}
  \label{alg:offline-learning}
  \small
  \begin{algorithmic}[1]
    \Require Program corpus \(\mathcal{P}\), search budget \(K\)
    \Ensure Skill library \(\mathcal{T}\)

    \State \(\mathcal{T} \gets \emptyset\)
    \ForAll{\(P \in \mathcal{P}\)}
    \LComment{Profile the original program.}
    \State \((\mathcal{E}_P, C_0) \gets \textsc{ProfileSE}(P)\)
    \If{\(\textsc{Saturated}(C_0)\)}
    \State \textbf{continue}
    \EndIf

    \LComment{Infer candidate barriers from code and SE traces.}
    \State \(\mathcal{B} \gets \textsc{ProposeBarriers}(P,\mathcal{E}_P)\)

    \ForAll{\(b \in \mathcal{B}\)}
    \LComment{Search transformed representations and validate by replay on \(P\).}
    \State \(\mathcal{V}_b \gets \textsc{SearchTransforms}(P,b,C_0,K)\)
    \If{\(\mathcal{V}_b \neq \emptyset\)}
    \LComment{Distill validated examples into reusable skills.}
    \State \(\mathcal{T} \gets \mathcal{T} \cup \textsc{CreateSkills}(b,\mathcal{V}_b)\)
    \EndIf
    \EndFor
    \EndFor

    \State \Return \(\mathcal{T}\)
  \end{algorithmic}
\end{algorithm}

\paragraph{Rule learning}
While validated transformation examples are useful in-context demonstrations for the agent, a single before/after pair is too specific to transfer directly to large projects (\Cref{sec:methodology/online-inference}).
To this end, \symbolon generalizes validated examples into agent skills.
For each barrier, the agent first synthesizes additional variants that express the same transformation idea under different local contexts.
Each variant is validated with the same transform-and-replay metric.
\symbolon then groups retained examples by barrier semantics and transformation strategy, deduplicates equivalent groups, and converts each group into a skill.

Specifically, a skill includes four components: the barrier description, the applicability context, the transformation procedure, and representative before/after examples with replay evidence (\Cref{app:transformation_table}).
For instance, a skill may describe when a byte-by-byte comparison can be summarized into a symbolic choice plus an assumption, together with examples such as \texttt{strcmp}, \texttt{memcmp}, or project-specific wrapper functions.
The resulting skill library \(\mathcal{T}\) can be further updated, \eg the rule-learning LLM can add, merge, and remove rules when it iterates through the samples in \(\mathcal{P}\), mimicking a typical training loop.
Importantly, \(\mathcal{T}\) is persistent and inspectable.
It transfers transformation knowledge from small programs to large projects without updating model weights.

\Cref{alg:offline-learning} abstracts the high-level steps.
\textsc{SearchTransforms} runs an LLM-guided evolutionary search.
Each candidate \(\widehat{P}\) is first checked for compilation, then evaluated by running symbolic execution on \(\widehat{P}\) and replaying the generated tests on the original program \(P\).
It returns validated examples \((P,\widehat{P})\) whose replayed metric improves over \(C_0\).

\subsection{Online Inference: Repo-Level Transformation}
\label{sec:methodology/online-inference}

Given a target project \(P_{\text{tgt}}\), \symbolon uses the learned skill library \(\mathcal{T}\) to produce a transformed project (scaffold) for symbolic execution.
The agent is given the current source tree, relevant skills retrieved based on its own judgment, and feedback from the previous valid iteration, including compiler diagnostics, symbolic-execution traces, and replayed coverage reports.
This makes the transformation context-sensitive in two ways.
First, the agent sees the local source context around each candidate edit, \eg control flow, variable names, types, and comments.
Second, the agent receives execution feedback that indicates which transformed regions helped or hindered exploration.
Therefore, a skill is not applied as a fixed textual rewrite, but is instantiated according to the local program representation.

In each iteration, the agent retrieves skills whose applicability conditions match the current code and proposes localized source edits.
\symbolon then validates the candidate in three steps.
First, it checks that the transformed project builds into LLVM bitcode.
Second, it runs symbolic execution on the transformed bitcode.
Third, it replays the generated tests on the original target project \(P_{\text{tgt}}\) and measures the replayed metric.
If compilation fails, the candidate is rejected, and the diagnostics are returned to the agent.
If compilation succeeds, the candidate becomes a valid scaffold, and \symbolon updates the selected output \(P_{\text{sel}}\) only when the replayed metric improves.
After at most \(K\) iterations, \symbolon returns \(P_{\text{sel}}\), the best validated scaffold observed under the transformation budget.
The original project remains the measurement target throughout this process.
The transformed scaffold is used only to generate tests, while replay on \(P_{\text{tgt}}\) determines whether a transformation is useful.
This design filters invalid or unhelpful transfers from the offline skill library and keeps the optimization objective aligned with symbolic-execution progress on the original code.

\begin{algorithm}[!t]
  \caption{Skill-guided repo-level transformation}
  \label{alg:online-inference}
  \small
  \begin{algorithmic}[1]
    \Require Target project \(P_{\text{tgt}}\), skill library \(\mathcal{T}\), transformation budget \(K\)
    \Ensure Selected transformed scaffold \(P_{\text{sel}}\)

    \State \(P_{\text{cur}} \gets P_{\text{tgt}}\), \(P_{\text{sel}} \gets P_{\text{tgt}}\)
    \LComment{Initial feedback and baseline metric are measured on the original target.}
    \State \((F, C_{\text{sel}}) \gets \textsc{Evaluate}(P_{\text{tgt}}, P_{\text{tgt}})\)

    \For{\(i = 1\) to \(K\)}
    \LComment{Agent uses source context, learned skills, and prior feedback.}
    \State \(\widehat{P} \gets \textsc{AgentTransform}(P_{\text{cur}},\mathcal{T},F)\)

    \If{\(\neg \textsc{Builds}(\widehat{P})\)}
    \State \(F \gets \textsc{Diagnostics}(\widehat{P})\)
    \State \textbf{continue}
    \EndIf

    \LComment{Run SE on the scaffold, replay tests on the original target.}
    \State \((F,C) \gets \textsc{Evaluate}(\widehat{P},P_{\text{tgt}})\)
    \State \(P_{\text{cur}} \gets \widehat{P}\)

    \If{\(C > C_{\text{sel}}\)}
    \State \(P_{\text{sel}} \gets \widehat{P}\), \(C_{\text{sel}} \gets C\)
    \EndIf
    \EndFor

    \State \Return \(P_{\text{sel}}\)
  \end{algorithmic}
\end{algorithm}

\Cref{alg:online-inference} summarizes this online inference step, which adapts learned agent skills for repo-level transformations.
\textsc{Evaluate}(\(\widehat{P}, P\)) runs symbolic execution on \(\widehat{P}\), replays the generated tests on \(P\), and returns both the replayed metric and feedback artifacts such as traces, coverage reports, and diagnostics.
In our evaluation, we use a fixed skill library learned offline.
A natural extension is to promote validated target-specific transformations back into the library, but we disable this online skill growth in our evaluation to keep costs manageable (we discuss this extension in \Cref{sec:discussion}).

\section{Implementation}
\label{sec:implementation}

We implement \symbolon in 2.1K LoC of Python, 3.5K LoC of shell scripts, and 3.7K LoC of Nix. We use KLEE~3.2 as the default symbolic execution backend.

\paragraph{Offline transformation learning corpus}
We construct the offline learning corpus from CodeContests\cite{li2022codecontests}, which contains millions of competitive-programming submissions.
To obtain C programs, we compile submissions with \texttt{gcc} and keep those that compile successfully.
This leads to 22{,}578 C programs.
We remove near-duplicates, \eg reformatted variants or versions with renamed identifiers, using a Jaccard-similarity threshold of 0.85, leaving 15{,}879 programs.
We then sample one program per problem, producing a final corpus of 2{,}416 programs with an average length of 50.5 lines.
Because these programs are small, we run KLEE on each with a 30-second timeout and an 8\,GB memory limit, using the command-line configuration shown in \Cref{app:klee-command}.

\paragraph{Transformation search}
We instantiate the rule-learning phase with OpenEvolve\cite{openevolve}, which uses an LLM to propose program variants and an external evaluator to select candidate transformed programs for refinement.
The evaluator runs KLEE on the transformed program, replays the generated tests on the original program, and uses the resulting coverage as the reward.
For each program, we cap the search at 20 iterations and stop early once replayed tests reach 100\% coverage on the original program.
Among the 2{,}416 programs, 1{,}567 reach 100\% coverage within 14 iterations, suggesting that many useful transformations are found within a small search budget.
We use Claude Sonnet~4.6 as the default model.
The offline rule-learning cost averages \$0.34 and 38.4K tokens per program.
This is distinct from the per-project deployment cost measured in \Cref{sec:evaluation/cost}.

\paragraph{Skill library}
The offline learning process produces a skill library with 211 skills after deduplication.
Each skill is stored as a directory that contains a \texttt{SKILL.md} file and a \texttt{references/} directory.
The \texttt{SKILL.md} file describes the code pattern, why such a pattern impedes symbolic execution, and the expected changes in symbolic execution traces after applying the transformation.
The \texttt{references/} directory stores validated transformation examples collected during offline learning.
On average, each skill contains 1.02 validated transformation examples and 2.3 augmented applicability patterns.
During online inference, the agent can retrieve a skill by recognizing a code pattern in the target program and instantiate the corresponding transformation according to the surrounding program context.

\paragraph{Agent settings}
We implement the repo-level transformation agent as a Python program built on Claude Agent SDK~0.2.96.
The learned skills are mounted as a local read-only plugin, and all skills are enabled during transformation.
Given a target project, the agent proposes localized source edits, while an external harness performs compilation, symbolic execution, and replay for coverage measurement.
To keep edits local, each round may modify at most three files.
After each round, the harness builds the transformed code, runs KLEE, collects its \texttt{run.istats} trace and an \texttt{llvm-cov} coverage report, and returns these artifacts to the agent for the next round.
The agent runs in a sandbox with write access only to the target codebase.
Transformation references and skills are read-only, network access and sub-agent spawning are disabled, and only a small whitelist of tools is available: read tools (\texttt{Read}, \texttt{Glob}, \texttt{Grep}, \texttt{Bash:sqlite3} for reading traces, and \texttt{Bash:jq} for reading coverage reports) and write tools (\texttt{Write}, \texttt{Edit}, and \texttt{MultiEdit}).
We guard symbolic-execution-only source edits with \texttt{\#ifdef KLEE\_BITCODE}, allowing the same source tree to produce a transformed LLVM-bitcode build for KLEE and an unmodified native build for replay-based coverage measurement.

\section{Evaluation}
\label{sec:evaluation}

To evaluate the effectiveness of \symbolon, we aim to answer the following
research questions:

\begin{rqlist}
  \rqitem[Effectiveness]\label{rq:coverage} How much does \symbolon help symbolic execution to improve code coverage with different search strategies?
  \rqitem[Execution Efficiency]\label{rq:overhead} How does \symbolon affect the runtime overhead of symbolic execution?
  \rqitem[Security Impact]\label{rq:realworld} How does \symbolon help find more real-world security violations and bugs?
  \rqitem[Cost]\label{rq:cost} What is the deployment cost of \symbolon?
  \rqitem[Sensitivity]\label{rq:ablation} How much do different components of \symbolon contribute to its overall effectiveness?
\end{rqlist}

\subsection{Experiment Setup}
\label{sec:evaluation/setup}

\begin{table}[!t]
  \centering
  \caption{Programs used to evaluate \symbolon. LoC is the number of source lines reported by \texttt{llvm-cov} and serves as the upper bound for line coverage.
  }
  \label{tab:program-stats}
  \small
  \resizebox{\columnwidth}{!}%
  {%
    \begin{tabular}{l l r | l l r}
      \toprule
      \textbf{Program} & \textbf{Version} & \textbf{LoC}
                       & \textbf{Program} & \textbf{Version} & \textbf{LoC} \\
      \midrule
      bash             & 5.3              & 55{,}638         & bc        & 1.08.2   & 5{,}339 \\
      bison            & 3.8.2            & 29{,}108         & cjpeg     & 3.1.4.1  & 8{,}642 \\
      cjson            & 1.7.19           & 2{,}298          & combine   & 0.4.0    & 5{,}332 \\
      curl             & 8.20.0           & 58{,}091         & diff      & 3.12     & 8{,}898 \\
      expat            & 2.8.0            & 9{,}711          & find      & 4.10.0   & 15{,}930 \\
      flex             & 2.6.4            & 9{,}451          & flvmeta   & 1.2.2    & 7{,}478 \\
      gawk             & 5.4.0            & 41{,}246         & grep      & 3.12     & 10{,}303 \\
      gzip             & 1.14             & 4{,}074          & jasper    & 4.2.9    & 22{,}124 \\
      jq               & 1.8.1            & 16{,}935         & xmllint   & 2.15.3   & 88{,}109 \\
      lua              & 5.4.7            & 14{,}883         & make      & 4.4      & 14{,}070 \\
      nasm             & 3.01             & 49{,}625         & objcopy   & 2.46.0   & 97{,}417 \\
      openssl          & 4.0.0            & 283{,}323        & patch     & 2.8      & 10{,}525 \\
      readelf          & 2.46.0           & 60{,}612         & sed       & 4.10     & 13{,}852 \\
      sqlite           & 3.53.1           & 116{,}524        & strip-new & 2.46.0   & 97{,}658 \\
      tic              & 6.6              & 11{,}994         & tiffinfo  & 4.7.1    & 19{,}725 \\
      transicc         & 2.19             & 19{,}742         & vim       & 9.2.0458 & 112{,}907 \\
      \bottomrule
    \end{tabular}%
  }
\end{table}

\paragraph{Benchmark}
We evaluate \symbolon on 32 real-world open-source programs widely used in previous symbolic execution and fuzzing studies \cite{yao2025empc, sun2024cgs, he2021learch, bohme2017aflgo, busse2020moklee, metzman2021fuzzbench}, following the same configurations when settings are available.
As shown in \Cref{tab:program-stats}, we use the latest release available for each program at the time of our evaluation.
The exact symbolic environment supplied to each program, \ie the concrete arguments and \texttt{--sym-*} options appended to each KLEE invocation, is listed in \Cref{app:program-symenv}.
We build all projects using \texttt{clang-16} and disable compiler optimization (\texttt{-O0}) to ensure the generated bitcode is consistent with the original code implementation.
Other build configurations include flags minimizing dependence on external libraries.

\paragraph{Baselines}
We evaluate \symbolon across a broad set of established search strategies to assess whether \symbolon-transformed programs are generally amenable to symbolic execution, rather than merely tailored to particular strategies.
Specifically, we include all 11 built-in search strategies in KLEE, together with five additional representative works from recent years, summarized in \Cref{tab:search-strategies}.
All experiments are built on KLEE~3.2\cite{cadar2008klee}.
Empc\cite{yao2025empc} runs with the original settings, including the bipartite-matching cap and analyzer thread-pool configuration.
CGS\cite{sun2024cgs} requires a branch-dependency pre-pass, which we apply to the bitcode before running it with the default target-branch selection settings.
We evaluate Learch\cite{he2021learch} with its shipped feedforward network under greedy state selection, which matches their reported best-performing configuration.
SGS\cite{li2013sgs} interleaves four subpath-guided searchers with subpath lengths fixed to 1/2/4/8 over each state's most recently taken branches.
For CBC\cite{yi2024cbc}, we use its default settings for the bounded-DFS exploration depth around sensitive instructions.

\paragraph{Metrics}
We use line coverage as the primary metric following prior work\cite{cadar2008klee,yao2025empc,poeplau2020symcc}, considering that \symbolon transforms the program before symbolic execution, making KLEE's internal bitcode coverage not representative of the coverage on the original source code.
We thus report external coverage: after KLEE generates tests on either the original or transformed code, we replay those tests on a separately instrumented native build of the original program and measure line coverage with \texttt{llvm-cov}.
The coverage is measured over 10 independent runs with a 12-hour end-to-end budget.
For \symbolon, this budget includes repo-level transformation and validation time, so any time spent by the agent directly reduces the remaining symbolic-execution time, ensuring a fair comparison with the baselines.

\begin{table}[!t]
  \centering
  \caption{Baseline search strategies used in our evaluation.
  }
  \label{tab:search-strategies}
  \small
  \begin{tabularx}{\columnwidth}{@{} l @{\hspace{2em}} >{\raggedright\arraybackslash}X @{}}
    \toprule
    \textbf{Strategy}                  & \textbf{Prioritization} \\
    \midrule
    \texttt{bfs}                       & Oldest live state, exploring shallow paths \\
    \texttt{dfs}                       & Most recently added state \\
    \texttt{random-path}               & Random on execution tree, prefer deep states \\
    \texttt{random-state}              & Any live path with uniform probability \\
    \texttt{nurs:depth}                & Deeper states with higher probability \\
    \texttt{nurs:covnew}               & Paths that recently discovered new coverage \\
    \texttt{nurs:cpicnt}               & States whose call-path is rarely taken \\
    \texttt{nurs:qc}                   & States with cheap constraint-solving history \\
    \texttt{nurs:icnt}                 & States at rarely-executed instructions \\
    \texttt{nurs:md2u}                 & Paths closest to uncovered CFG instructions \\
    \texttt{nurs:rp}                   & Exponentially favors shallower states \\
    \midrule
    \texttt{cbc}\cite{yi2024cbc}       & Reviving states near sensitive instructions \\
    \texttt{cgs}\cite{sun2024cgs}      & Paths storing flip uncovered branches \\
    \texttt{empc}\cite{yao2025empc}    & Paths in minimum path cover of the CFG \\
    \texttt{learch}\cite{he2021learch} & Highest-reward state from trained model \\
    \texttt{sgs}\cite{li2013sgs}       & The least explored subpaths \\
    \bottomrule
  \end{tabularx}
\end{table}

\paragraph{Environment}
We run all evaluations on bare-metal x86-64 servers, each with two Intel Xeon Gold 6240R processors and 192\,GB RAM, running Ubuntu~24.04.4 LTS \cite{keahey2020lessons}.
Each evaluation instance runs in an isolated container limited to one CPU core and 40\,GB memory.
We use Z3\cite{moura2008z3} as the SMT backend and cap KLEE at 32\,GB memory, leaving the remaining container memory for the solver, runtime libraries, and OS overhead.
All runs use the same fixed KLEE configuration, including POSIX/uClibc runtime support, query logging, coverage output, and a watchdog (see the full command line in \Cref{app:klee-command}).
To reduce resource interference, we run at most four evaluation instances concurrently on each server and disable other background workloads.

\subsection{Coverage Improvement}
\label{sec:evaluation/rq1}
\input{tex/figures/coverage}

\Cref{fig:coverage} summarizes the coverage improvement of \symbolon across 32 programs and 16 search strategies.
\symbolon substantially improves replayed line coverage on the original programs.
The average improvement is 3.69\(\times\) and 3.87\(\times\) when coverage is aggregated by programs and search strategies, respectively.
The gains are especially visible in hard cases where the baseline makes no progress.
For 69 program--strategy pairs with zero baseline coverage, \symbolon covers at least one line, and for 68 of these pairs, it covers at least 1{,}000 lines.

The gains hold across most benchmark programs.
For each program, we compute the improvement ratio by summing the covered lines over all 16 search strategies with \symbolon and dividing by the corresponding baseline sum.
\symbolon improves coverage on 30 of the 32 programs, with at least 1.5\(\times\) improvement on 25 programs.
On 19 programs, \symbolon covers more lines than the baseline under every evaluated search strategy.
The largest program-level improvement is 22.75\(\times\) on \texttt{lua}.
We leave the discussion of why \symbolon does not improve \texttt{cjson} or \texttt{nasm} to \Cref{sec:discussion}, though it still improves a subset of search strategies.

\symbolon also improves coverage across all search strategies.
As some programs have zero baseline coverage, improvement ratios for baselines on these programs are undefined.
Therefore, for each strategy, we report an aggregate improvement ratio, computed as the total number of lines covered by \symbolon across all 32 programs divided by the corresponding total baseline coverage.
The improvement ranges from 1.56\(\times\) to 7.51\(\times\), with a mean of 3.87\(\times\).
Notably, \symbolon also reduces the dependence on the choice of search strategy.
On the original programs, the strongest strategy covers 7.4\(\times\) as many lines as the weakest one.
After applying \symbolon, this gap drops to 1.7\(\times\).
This suggests that \symbolon makes the transformed programs easier to explore across a wide range of search strategies, rather than only benefiting a particular strategy.

\definecolor{cfr0}{RGB}{31,119,180}
\definecolor{cfb0}{RGB}{143,187,218}
\definecolor{cfr1}{RGB}{255,127,14}
\definecolor{cfb1}{RGB}{255,191,134}
\definecolor{cfr2}{RGB}{44,160,44}
\definecolor{cfb2}{RGB}{150,208,150}
\definecolor{cfr3}{RGB}{214,39,40}
\definecolor{cfb3}{RGB}{234,147,148}
\definecolor{cfr4}{RGB}{148,103,189}
\definecolor{cfb4}{RGB}{202,179,222}
\definecolor{cfr5}{RGB}{140,86,75}
\definecolor{cfb5}{RGB}{198,170,165}
\definecolor{cfr6}{RGB}{227,119,194}
\definecolor{cfb6}{RGB}{241,187,224}
\pgfplotsset{
  covcflagv axis/.style={
    scale only axis, width=2.50cm, height=1.60cm,
    ymin=0, enlarge x limits=false, xmin=-0.7, xmax=6.7,
    xtick=\empty, ytick=\empty, tick style={draw=none},
    axis line style={black, line width=0.4pt}, axis on top,
  },
}
\begin{figure}[!t]
  \centering
  \resizebox{0.88\linewidth}{!}{%
    \begin{tikzpicture}[
        rowlbl/.style={font=\footnotesize, anchor=east, inner sep=2pt},
        nm/.style={font=\footnotesize, anchor=north, align=center, inner sep=1.6pt},
      ]
      \node[rowlbl] at (-0.28,0.48) {\(+\)\,\symbolon};
      \node[rowlbl] at (-0.28,0.14) {Baseline};
      \filldraw[fill=cfr0, draw=black, line width=0.2pt] (0.00,0.00) rectangle (1.00,0.28);
      \filldraw[fill=cfb0, draw=black, line width=0.2pt] (0.00,0.34) rectangle (1.00,0.62);
      \filldraw[fill=cfr1, draw=black, line width=0.2pt] (1.68,0.00) rectangle (2.68,0.28);
      \filldraw[fill=cfb1, draw=black, line width=0.2pt] (1.68,0.34) rectangle (2.68,0.62);
      \filldraw[fill=cfr2, draw=black, line width=0.2pt] (3.36,0.00) rectangle (4.36,0.28);
      \filldraw[fill=cfb2, draw=black, line width=0.2pt] (3.36,0.34) rectangle (4.36,0.62);
      \filldraw[fill=cfr3, draw=black, line width=0.2pt] (5.04,0.00) rectangle (6.04,0.28);
      \filldraw[fill=cfb3, draw=black, line width=0.2pt] (5.04,0.34) rectangle (6.04,0.62);
      \filldraw[fill=cfr4, draw=black, line width=0.2pt] (6.72,0.00) rectangle (7.72,0.28);
      \filldraw[fill=cfb4, draw=black, line width=0.2pt] (6.72,0.34) rectangle (7.72,0.62);
      \filldraw[fill=cfr5, draw=black, line width=0.2pt] (8.40,0.00) rectangle (9.40,0.28);
      \filldraw[fill=cfb5, draw=black, line width=0.2pt] (8.40,0.34) rectangle (9.40,0.62);
      \filldraw[fill=cfr6, draw=black, line width=0.2pt] (10.08,0.00) rectangle (11.08,0.28);
      \filldraw[fill=cfb6, draw=black, line width=0.2pt] (10.08,0.34) rectangle (11.08,0.62);
      \node[nm] at (0.50,-0.06) {\strut vanilla\\\strut (3.39\(\times\))};
      \node[nm] at (2.18,-0.06) {\strut \texttt{-O2}\\\strut (3.77\(\times\))};
      \node[nm] at (3.86,-0.06) {\strut \texttt{-sccp}\\\strut (3.21\(\times\))};
      \node[nm] at (5.54,-0.06) {\strut \texttt{-ipsccp}\\\strut (3.69\(\times\))};
      \node[nm] at (7.22,-0.06) {\strut \texttt{-gvn\_hoist}\\\strut (3.44\(\times\))};
      \node[nm] at (8.90,-0.06) {\strut \texttt{-sroa}\\\strut (3.34\(\times\))};
      \node[nm] at (10.58,-0.06) {\strut \texttt{-adce}\\\strut (3.77\(\times\))};
  \end{tikzpicture}}
  \par\vspace{4pt}
  {%
    \begin{tikzpicture}
      \begin{groupplot}[
          group style={group size=2 by 1, horizontal sep=6pt},
          covcflagv axis,
        ]
        \nextgroupplot[ymax=13195]
        \filldraw[fill=cfr0, draw=black, line width=0.2pt] (axis cs:-0.42,0) rectangle (axis cs:0.42,2098);
        \filldraw[fill=cfb0, draw=black, line width=0.2pt] (axis cs:-0.42,2098) rectangle (axis cs:0.42,10695);
        \filldraw[fill=cfr1, draw=black, line width=0.2pt] (axis cs:0.58,0) rectangle (axis cs:1.42,2005);
        \filldraw[fill=cfb1, draw=black, line width=0.2pt] (axis cs:0.58,2005) rectangle (axis cs:1.42,10995);
        \filldraw[fill=cfr2, draw=black, line width=0.2pt] (axis cs:1.58,0) rectangle (axis cs:2.42,2195);
        \filldraw[fill=cfb2, draw=black, line width=0.2pt] (axis cs:1.58,2195) rectangle (axis cs:2.42,10636);
        \filldraw[fill=cfr3, draw=black, line width=0.2pt] (axis cs:2.58,0) rectangle (axis cs:3.42,1998);
        \filldraw[fill=cfb3, draw=black, line width=0.2pt] (axis cs:2.58,1998) rectangle (axis cs:3.42,10995);
        \filldraw[fill=cfr4, draw=black, line width=0.2pt] (axis cs:3.58,0) rectangle (axis cs:4.42,2005);
        \filldraw[fill=cfb4, draw=black, line width=0.2pt] (axis cs:3.58,2005) rectangle (axis cs:4.42,10995);
        \filldraw[fill=cfr5, draw=black, line width=0.2pt] (axis cs:4.58,0) rectangle (axis cs:5.42,2004);
        \filldraw[fill=cfb5, draw=black, line width=0.2pt] (axis cs:4.58,2004) rectangle (axis cs:5.42,10230);
        \filldraw[fill=cfr6, draw=black, line width=0.2pt] (axis cs:5.58,0) rectangle (axis cs:6.42,1997);
        \filldraw[fill=cfb6, draw=black, line width=0.2pt] (axis cs:5.58,1997) rectangle (axis cs:6.42,10995);
        \node[anchor=north, font=\tiny, inner sep=1.6pt] at (rel axis cs:0.5,1) {jasper 5.28\(\times\)};
        \nextgroupplot[ymax=4507]
        \filldraw[fill=cfr0, draw=black, line width=0.2pt] (axis cs:-0.42,0) rectangle (axis cs:0.42,2103);
        \filldraw[fill=cfb0, draw=black, line width=0.2pt] (axis cs:-0.42,2103) rectangle (axis cs:0.42,3539);
        \filldraw[fill=cfr1, draw=black, line width=0.2pt] (axis cs:0.58,0) rectangle (axis cs:1.42,1903);
        \filldraw[fill=cfb1, draw=black, line width=0.2pt] (axis cs:0.58,1903) rectangle (axis cs:1.42,3755);
        \filldraw[fill=cfr2, draw=black, line width=0.2pt] (axis cs:1.58,0) rectangle (axis cs:2.42,2218);
        \filldraw[fill=cfb2, draw=black, line width=0.2pt] (axis cs:1.58,2218) rectangle (axis cs:2.42,3539);
        \filldraw[fill=cfr3, draw=black, line width=0.2pt] (axis cs:2.58,0) rectangle (axis cs:3.42,1944);
        \filldraw[fill=cfb3, draw=black, line width=0.2pt] (axis cs:2.58,1944) rectangle (axis cs:3.42,3537);
        \filldraw[fill=cfr4, draw=black, line width=0.2pt] (axis cs:3.58,0) rectangle (axis cs:4.42,2213);
        \filldraw[fill=cfb4, draw=black, line width=0.2pt] (axis cs:3.58,2213) rectangle (axis cs:4.42,3523);
        \filldraw[fill=cfr5, draw=black, line width=0.2pt] (axis cs:4.58,0) rectangle (axis cs:5.42,2109);
        \filldraw[fill=cfb5, draw=black, line width=0.2pt] (axis cs:4.58,2109) rectangle (axis cs:5.42,3518);
        \filldraw[fill=cfr6, draw=black, line width=0.2pt] (axis cs:5.58,0) rectangle (axis cs:6.42,1903);
        \filldraw[fill=cfb6, draw=black, line width=0.2pt] (axis cs:5.58,1903) rectangle (axis cs:6.42,3725);
        \node[anchor=north, font=\tiny, inner sep=1.6pt] at (rel axis cs:0.5,1) {make 1.75\(\times\)};
      \end{groupplot}
  \end{tikzpicture}}
  \caption{Line coverage of \symbolon under different compiler-optimization
    flags. Each bar is one flag (color); the solid base is the baseline coverage
    with that flag and the lighter cap is the additional lines \symbolon covers.
  Titles report the per-program improvement ratio.}
  \label{fig:coverage-compiler}
\end{figure}
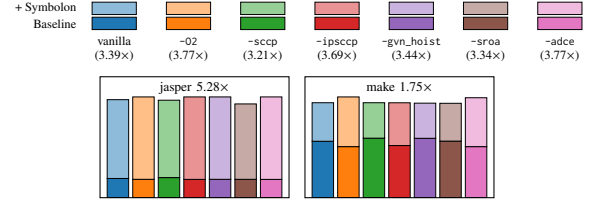

\paragraph{Improving compiler optimizations}
Beyond improving search strategies, we show that \symbolon also complements existing transformation-based approaches, \eg compiler optimizations\cite{zhang2024compiler}.
Specifically, for a program compiled with a compiler optimization flag deemed effective by prior work\cite{zhang2024compiler}, we further transform it with \symbolon and compare the resulting coverage improvement against that on the original program.
As shown in \Cref{fig:coverage-compiler}, \symbolon consistently improves coverage across all compiler-flag configurations by 3.52\(\times\) on average.
This observation validates that \symbolon can further augment rigid compiler transformations, which are strictly semantics-preserving and often misaligned with the objective of symbolic execution.

\subsection{Execution Efficiency}
\label{sec:evaluation/rq2}

\begin{table}[!t]
  \centering
  \caption{Runtime overhead of symbolically executing the original
    (Vanilla) versus the \symbolon-transformed program, for peak resident
    memory and average solving time per query. The reduction factor \emph{Ratio} column is computed as Vanilla/\symbolon (a ratio \({>}1\)
    means \symbolon\ costs less and higher is better); the last row reports arithmetic means across
    programs.
  }
  \label{tab:runtime-overhead}
  \small
  \resizebox{\columnwidth}{!}%
  {%
    \begin{tabular}{l r r r r r r}
      \toprule
      \multirow{2}[2]{*}{\textbf{Program}}
                         & \multicolumn{3}{c}{\textbf{Peak Memory (MB)}}
                         & \multicolumn{3}{c}{\textbf{Avg. Solving Time (ms)}} \\
      \cmidrule(lr){2-4} \cmidrule(lr){5-7}
                         & Vanilla    & \symbolon & Ratio
                         & Vanilla    & \symbolon & Ratio \\
      \midrule
      \texttt{bash}      & 29{,}998   & 14{,}215  & \gc{22}2.1  & 5{,}543    & 983       & \gc{32}5.6 \\
      \texttt{bc}        & 19{,}353   & 260       & \gc{62}74.3 & 1{,}739    & 5.7       & \gc{70}307 \\
      \texttt{bison}     & 7{,}751    & 1{,}028   & \gc{32}7.5  & 16{,}601   & 90        & \gc{70}184 \\
      \texttt{cjpeg}     & 19{,}279   & 16{,}362  & \gc{12}1.2  & 41{,}427   & 13{,}703  & \gc{22}3.0 \\
      \texttt{cjson}     & 26{,}845   & 24{,}189  & \gc{12}1.1  & 282        & 260       & \gc{12}1.1 \\
      \texttt{combine}   & 26{,}902   & 405       & \gc{62}66.4 & 43         & 14        & \gc{22}3.2 \\
      \texttt{curl}      & 17{,}719   & 292       & \gc{62}60.6 & 318        & 2.4       & \gc{70}133 \\
      \texttt{diff}      & 14{,}116   & 1{,}727   & \gc{32}8.2  & 18{,}854   & 212       & \gc{62}88.9 \\
      \texttt{expat}     & 15{,}625   & 305       & \gc{62}51.3 & 4{,}577    & 26        & \gc{70}178 \\
      \texttt{find}      & 22{,}211   & 616       & \gc{52}36.1 & 486        & 36        & \gc{42}13.5 \\
      \texttt{flex}      & 96         & 186       & 0.52        & 1.5        & 5.8       & 0.25 \\
      \texttt{flvmeta}   & 25{,}644   & 336       & \gc{62}76.4 & 2{,}855    & 46        & \gc{62}61.6 \\
      \texttt{gawk}      & 23{,}464   & 25{,}194  & 0.93        & 6{,}232    & 16        & \gc{70}384 \\
      \texttt{grep}      & 13{,}194   & 11{,}683  & \gc{12}1.1  & 35{,}440   & 8{,}836   & \gc{22}4.0 \\
      \texttt{gzip}      & 1{,}142    & 994       & \gc{12}1.1  & 192{,}424  & 4{,}211   & \gc{52}45.7 \\
      \texttt{jasper}    & 19{,}174   & 205       & \gc{62}93.6 & 343        & 2.3       & \gc{70}149 \\
      \texttt{jq}        & 22{,}203   & 283       & \gc{62}78.5 & 27         & 1{,}195   & 0.02 \\
      \texttt{xmllint}   & 17{,}604   & 444       & \gc{52}39.7 & 5{,}011    & 31        & \gc{70}159 \\
      \texttt{lua}       & 11{,}354   & 5{,}981   & \gc{12}1.9  & 215        & 1{,}918   & 0.11 \\
      \texttt{make}      & 23{,}199   & 19{,}377  & \gc{12}1.2  & 6{,}289    & 1{,}189   & \gc{32}5.3 \\
      \texttt{nasm}      & 22{,}011   & 26{,}941  & 0.82        & 10{,}405   & 87        & \gc{70}120 \\
      \texttt{objcopy}   & 18{,}916   & 1{,}726   & \gc{42}11.0 & 8{,}196    & 4{,}824   & \gc{12}1.7 \\
      \texttt{openssl}   & 16{,}969   & 2{,}309   & \gc{32}7.3  & 12{,}396   & 36        & \gc{70}342 \\
      \texttt{patch}     & 32{,}094   & 28{,}756  & \gc{12}1.1  & 659        & 570       & \gc{12}1.2 \\
      \texttt{readelf}   & 649        & 271       & \gc{22}2.4  & 53{,}188   & 166       & \gc{70}321 \\
      \texttt{sed}       & 26{,}368   & 26{,}734  & 0.99        & 408        & 249       & \gc{12}1.6 \\
      \texttt{sqlite}    & 26{,}356   & 768       & \gc{52}34.3 & 4{,}975    & 13        & \gc{70}397 \\
      \texttt{strip-new} & 18{,}891   & 462       & \gc{52}40.8 & 8{,}079    & 42        & \gc{70}193 \\
      \texttt{tic}       & 25{,}562   & 308       & \gc{62}82.9 & 348        & 0.99      & \gc{70}351 \\
      \texttt{tiffinfo}  & 14{,}884   & 326       & \gc{52}45.6 & 725        & 6.5       & \gc{70}111 \\
      \texttt{transicc}  & 25{,}815   & 317       & \gc{62}81.4 & 2{,}821    & 18        & \gc{70}155 \\
      \texttt{vim}       & 19{,}094   & 881       & \gc{42}21.7 & 143{,}495  & 624       & \gc{70}230 \\
      \midrule
      \textbf{Average}   & {18{,}890} & {6{,}684} & {29.2}      & {18{,}263} & {1{,}232} & {123} \\
      \bottomrule
    \end{tabular}%
  }
\end{table}

Symbolic execution can stop making progress when the state set or solver workload exhausts the analysis budget.
We thus evaluate whether \symbolon makes programs cheaper to explore.
For each program, we report peak resident memory and average solver time per query, averaged across all 16 search strategies.
As shown in \Cref{tab:runtime-overhead}, \symbolon reduces peak memory on 28 of 32 programs and average solver time per query on 29 of 32 programs.
The mean per-program reduction factors are 29.2\(\times\) for peak memory and 123\(\times\) for solver time per query.

\paragraph{Memory}
\symbolon reduces memory by removing or compressing sources of state growth before the engine reaches them, \eg summarizing byte-by-byte checks, bounding symbolic loops, and bypassing exploration-irrelevant I/O.
The arithmetic mean peak memory drops from 18{,}890\,MB to 6{,}684\,MB, while the median transformed peak memory is only 824\,MB.
Moreover, 18 transformed programs run under 1\,GB.
The largest reductions exceed 70\(\times\) on memory-bound programs such as \texttt{jasper}, \texttt{tic}, and \texttt{bc}, indicating that \symbolon substantially reduces the live state space that KLEE must keep resident.

\paragraph{Constraint solving time}
\symbolon also reduces solver-side cost by changing the expressions that enter solver queries, \eg concretizing symbolic sizes or bounds and rewriting solver-hostile computations.
The arithmetic mean solver time per query drops from 18{,}263\,ms to 1{,}232\,ms, and the mean per-program reduction factor is 123\(\times\).
The largest reductions exceed 300\(\times\) on query-bound programs such as \texttt{sqlite}, \texttt{gawk}, and \texttt{openssl}.
These savings show that \symbolon improves exploration not only by reaching more code, but also by making transformed programs cheaper for both the engine and the solver to process.

\subsection{Finding Bugs}
\label{sec:evaluation/rq3}
We evaluate the ability of \symbolon to help discover real-world security violations.
Specifically, we apply \symbolon to (1) directly find security violations in user-space programs, and (2) generate syscall descriptions using SyzSpec\cite{hao2025syzspec} for the Linux kernel to assist kernel fuzzing.

\paragraph{Finding security violations with sanitizers}
We first use KLEE and \symbolon to detect memory-safety bugs and undefined behaviors in the same set of programs.
This leverages KLEE's built-in memory error detector and its integration with the Undefined Behavior Sanitizer (UBSan).
Since the program is transformed by \symbolon, sanitizer reports obtained from the transformed program do not necessarily indicate violations in the original program.
Following similar practices in prior work\cite{yao2025empc, sun2024cgs}, we build the original program with UBSan and Address Sanitizer (ASan), replay the tests generated by KLEE/\symbolon on it, and count the unique security violations reported by each sanitizer.

As \Cref{tab:security-violations} shows, across 16 search strategies, \symbolon reports 240 more UBSan violations and 25 more ASan violations than running KLEE on the original programs.
We have manually validated these sanitizer reports and have responsibly disclosed confirmed issues that present real security risks to the respective maintainers.

\begin{table}[!t]
  \centering
  \caption{
    Unique security violations found in original programs.
    Tests are generated on original (Vanilla) versus \symbolon-transformed programs.
    \(\Delta\) is the absolute gain achieved by \symbolon.
  }
  \label{tab:security-violations}
  \small
  \resizebox{\columnwidth}{!}%
  {%
    \begin{tabular}{l rrr rrr}
      \toprule
      \multirow{2}[2]{*}{\textbf{Strategy}}
                            & \multicolumn{3}{c}{\textbf{UBSan Violations}}
                            & \multicolumn{3}{c}{\textbf{ASan Violations}} \\
      \cmidrule(lr){2-4} \cmidrule(lr){5-7}
                            & Vanilla & \symbolon & \textbf{\(\Delta\)} & Vanilla & \symbolon & \textbf{\(\Delta\)} \\
      \midrule
      \texttt{bfs}          & 26      & 47        & \gc{48}+21          & 5       & 5         & 0 \\
      \texttt{dfs}          & 13      & 24        & \gc{29}+11          & 2       & 1         & -1 \\
      \texttt{random-path}  & 21      & 43        & \gc{49}+22          & 3       & 5         & \gc{26}+2 \\
      \texttt{random-state} & 14      & 19        & \gc{18}+5           & 6       & 9         & \gc{41}+3 \\
      \texttt{nurs:depth}   & 16      & 26        & \gc{27}+10          & 0       & 1         & \gc{12}+1 \\
      \texttt{nurs:covnew}  & 17      & 42        & \gc{55}+25          & 2       & 5         & \gc{41}+3 \\
      \texttt{nurs:cpicnt}  & 24      & 39        & \gc{36}+15          & 3       & 2         & -1 \\
      \texttt{nurs:qc}      & 14      & 19        & \gc{18}+5           & 2       & 1         & -1 \\
      \texttt{nurs:icnt}    & 15      & 24        & \gc{25}+9           & 0       & 3         & \gc{41}+3 \\
      \texttt{nurs:md2u}    & 18      & 31        & \gc{33}+13          & 2       & 4         & \gc{26}+2 \\
      \texttt{nurs:rp}      & 29      & 45        & \gc{38}+16          & 3       & 8         & \gc{70}+5 \\
      \texttt{cbc}          & 20      & 31        & \gc{29}+11          & 4       & 7         & \gc{41}+3 \\
      \texttt{cgs}          & 32      & 44        & \gc{31}+12          & 6       & 10        & \gc{56}+4 \\
      \texttt{empc}         & 37      & 67        & \gc{64}+30          & 9       & 9         & 0 \\
      \texttt{learch}       & 29      & 62        & \gc{70}+33          & 6       & 7         & \gc{12}+1 \\
      \texttt{sgs}          & 8       & 10        & \gc{12}+2           & 1       & 2         & \gc{12}+1 \\
      \midrule
      \textbf{Total}        & {333}   & {573}     & {+240}              & {54}    & {79}      & {+25} \\
      \bottomrule
    \end{tabular}%
  }
\end{table}

\paragraph{Finding crashes by generating syscall descriptions}
Symbolic execution can support kernel fuzzing by generating syscall descriptions that help construct valid syscall inputs.
For example, SyzSpec\cite{hao2025syzspec} symbolically analyzes kernel syscall handlers and extracts input types, sizes, and constraints for \texttt{syzkaller}.
However, native kernel code is difficult for symbolic execution to analyze, so we use \symbolon as a front end to SyzSpec.
We run SyzSpec on both the original kernel and the \symbolon-transformed kernel, generate syscall descriptions from each, and then fuzz the unmodified original kernel with \texttt{syzkaller}.
Following SyzSpec's setup, we target the same Linux drivers and subsystems that remain available in our evaluated kernel version, excluding \texttt{capi20} because it has been removed.
Each target is fuzzed for 24 hours over three runs, using four CPU cores split across two QEMU instances.
We report average final KCOV edge coverage and the number of unique crashes.

\begin{table}[!t]
  \centering
  \caption{
    24-hour \texttt{syzkaller} fuzzing results using syscall descriptions generated by running SyzSpec on the original kernel (Vanilla) and on the \symbolon-transformed kernel.
  }
  \label{tab:syzspec}
  \small
  \resizebox{\columnwidth}{!}%
  {%
    \begin{tabular}{l rrr cc}
      \toprule
      \multirow{2}[2]{*}{\textbf{Kernel Target}}
                               & \multicolumn{3}{c}{\textbf{Edge Coverage}}
                               & \multicolumn{2}{c}{\textbf{Crashes}} \\
      \cmidrule(lr){2-4} \cmidrule(lr){5-6}
                               & Vanilla & \symbolon & Ratio        & Vanilla & \symbolon \\
      \midrule
      \texttt{i2c}             & 1,621   & 1,743     & \gc{13}1.08  & 2       & 2 \\
      \texttt{infiniband}      & 714     & 1,831     & \gc{29}2.56  & 0       & \gc{12}1 \\
      \texttt{input\_event}    & 1,940   & 1,798     & 0.93         & 1       & 1 \\
      \texttt{io\_uring}       & 1,708   & 2,727     & \gc{20}1.60  & 1       & 1 \\
      \texttt{mapper\_control} & 1,185   & 26,234    & \gc{70}22.14 & 1       & \gc{70}4 \\
      \texttt{ppp}             & 3,878   & 3,022     & 0.78         & 1       & \gc{12}2 \\
      \texttt{ptmx}            & 1,846   & 5,055     & \gc{30}2.74  & 1       & \gc{41}3 \\
      \texttt{sg}              & 3,044   & 2,844     & 0.93         & 4       & 1 \\
      \texttt{snd\_seq}        & 1,620   & 2,108     & \gc{16}1.30  & 0       & \gc{12}1 \\
      \texttt{uinput}          & 2,767   & 2,898     & \gc{12}1.05  & 1       & \gc{12}2 \\
      \midrule
      \textbf{Total}           & 20,323  & 50,260    & 2.47         & 12      & 18 \\
      \bottomrule
    \end{tabular}%
  }
\end{table}

\begin{table}[t]
  \centering
  \caption{New linux kernel bugs discovered by \symbolon.
    {File}: enclosing source file (kernel-relative).
    {Function}: enclosing kernel function.
  }
  \label{tab:linux-bugs}
  \small
  \resizebox{\columnwidth}{!}%
  {%
    \begin{tabular}{l l l}
      \toprule
      \textbf{File}                    & \textbf{Function}                          & \textbf{Type} \\
      \midrule
      \texttt{jfs/jfs\_imap.c}         & \texttt{diRead}                            & NPD \\
      \texttt{net/socket.c}            & \texttt{kernel\_sock\_shutdown}            & NPD \\
      \texttt{vidtv/vidtv\_psi.c}      & \texttt{vidtv\_psi\_ts\_psi...}\(^1\)      & NPD \\
      \texttt{jfs/file.c}              & \texttt{jfs\_open}                         & OOB \\
      \texttt{jfs/jfs\_txnmgr.c}       & \texttt{txAbort}                           & Assertion \\
      \texttt{ipv4/udp\_tunnel\_nic.c} & \texttt{udp\_tunnel\_nic\_device...}\(^2\) & UAF \\
      \texttt{gfs2/rgrp.c}             & \texttt{read\_rindex\_entry}               & WARN \\
      \texttt{hfsplus/super.c}         & \texttt{hfsplus\_commit\_superblock}       & Hang \\
      \texttt{events/core.c}           & \texttt{perf\_event\_release\_kernel}      & Hang \\
      \texttt{ipv4/ip\_input.c}        & \texttt{ip\_rcv}                           & RCU stall \\
      \texttt{mm/memfd.c}              & \texttt{memfd\_create}                     & RCU stall \\
      \texttt{fs/stat.c}               & \texttt{newfstat}                          & RCU stall \\
      \texttt{fs/stat.c}               & \texttt{newlstat}                          & RCU stall \\
      \texttt{fs/fcntl.c}              & \texttt{sys\_fcntl}                        & RCU stall \\
      \texttt{events/core.c}           & \texttt{perf\_fasync}                      & RCU stall \\
      \texttt{events/core.c}           & \texttt{perf\_release}                     & RCU stall \\
      \texttt{block/blk-mq.c}          & \texttt{blk\_mq\_free\_rqs}                & OOB \\
      \texttt{v4l2-core/v4l2-dev.c}    & \texttt{v4l2\_open}                        & UAF \\
      \texttt{module/main.c}           & \texttt{try\_module\_get}                  & OOB \\
      \texttt{jfs/jfs\_logmgr.c}       & \texttt{lbmIODone}                         & UAF \\
      \texttt{dvb-core/dvbdev.c}       & \texttt{dvb\_device\_put}                  & UAF \\
      \bottomrule
      \multicolumn{3}{l}{\scriptsize \(^1\)\texttt{vidtv\_psi\_ts\_psi\_write\_into}\ \ \ \ \ \ \ \ \ \ \ \ \ \ \ \(^2\)\texttt{udp\_tunnel\_nic\_device\_sync\_work}}

    \end{tabular}%
  }
\end{table}

\Cref{tab:syzspec} shows that descriptions generated from the \symbolon-transformed kernel improve end-to-end fuzzing effectiveness.
Across all targets, aggregate edge coverage increases from 20{,}323 to 50{,}260, a 2.47\(\times\) improvement over descriptions generated from the original kernel.
\symbolon improves edge coverage on 7 of 10 targets, with the largest gain on \texttt{mapper\_control}.
The generated descriptions also increase the total number of unique crashes from 12 to 18, improving crash discovery on 6 of 10 targets.
These results show that transforming the symbolic-analysis target can help SyzSpec recover more useful syscall descriptions and translate into better kernel-fuzzing outcomes on the original kernel.

We further run an extended continuous fuzzing campaign beyond the 24-hour comparison.
This campaign uncovers 21 previously unknown Linux kernel bugs that were not reported by \texttt{syzbot} at the time of discovery.
As shown in \Cref{tab:linux-bugs}, these bugs include use-after-free, out-of-bounds access, null-pointer dereference, hang, and RCU-stall cases.
At the time of writing, we have responsibly reported all identified bugs to the relevant maintainers and have suggested potential patches to help the fixing process.
Among these bugs, two of them have been acknowledged by the maintainers and one bug has been forwarded to the public mail list for discussion.
We continue to actively monitor the status of these reports and will provide update for each bug.

\begin{figure*}[t]
  \centering
  \begin{subfigure}[t]{0.49\textwidth}
    \centering
    \ifusepdfcode\codepdf{bug-flvmeta.pdf}\else
    \begin{codebox}[ghtiny, /tcb/.cd, minted options={fontsize=\scriptsize,breaklines,tabsize=4,autogobble,numbers=left,numbersep=3pt}, left=4mm, colframe=black, equal height group=symbolonbugbox, valign=bottom, top=0pt, bottom=0pt]{assets/code/flvmeta-stack-overflow.c}
    \end{codebox}\fi
    \caption{A buffer overflow bug in \texttt{yaml\_on\_tag}: a 64-bit tag offset is printed with \texttt{sprintf} into a 20-byte stack buffer, but a crafted FLV forces an offset whose 20-digit decimal form needs 21 bytes, writing one past \texttt{buffer[20]}.}
    \label{fig:bug-flvmeta}
  \end{subfigure}
  \hfill
  \begin{subfigure}[t]{0.49\textwidth}
    \centering
    \ifusepdfcode\codepdf{bug-jfs.pdf}\else
    \begin{codebox}[ghtiny, /tcb/.cd, minted options={fontsize=\scriptsize,breaklines,tabsize=4,autogobble,numbers=left,numbersep=3pt}, left=4mm, colframe=black, equal height group=symbolonbugbox, valign=bottom, top=0pt, bottom=0pt]{assets/code/oob-jfs-open.c}
    \end{codebox}\fi
    \caption{An out-of-bounds access in \texttt{jfs\_open}: \texttt{BLKTOAG} decodes an on-disk PXD without validating that the result indexes within \texttt{db\_active[MAXAG]}, so a crafted inode forces a negative index and an out-of-bounds atomic increment.}
    \label{fig:bug-jfs}
  \end{subfigure}
  \caption{
    Two representative bugs exposed by \symbolon-generated tests and missed by the evaluated baselines under the same budget.
    (a) An ASan-reported stack overflow in \texttt{flvmeta}.
    (b) An out-of-bounds access in Linux JFS that can corrupt kernel memory or trigger a crash.
  }
  \label{fig:bugs}
\end{figure*}

\paragraph{Case studies}
\Cref{fig:bugs} shows two representative bugs exposed by \symbolon-generated tests.
In \texttt{flvmeta}, \symbolon enables KLEE to reach \texttt{yaml\_on\_tag}, where a 64-bit tag offset is printed with \texttt{sprintf} into a 20-byte stack buffer.
The generated input makes the decimal offset require 21 bytes including the null terminator, triggering a one-byte ASan-reported stack overflow.
In the Linux JFS case, \symbolon helps SyzSpec recover syscall descriptions that guide \texttt{syzkaller} to \texttt{jfs\_open}, where an on-disk extent descriptor is decoded into an unchecked allocation-group index and used to access \texttt{db\_active[MAXAG]}.

\subsection{Deployment Cost}
\label{sec:evaluation/cost}

\begin{table}[!t]
  \centering
  \caption{Per-program deployment cost of \symbolon: wall-clock time, token usage, and dollar cost of applying the distilled skill library with default model configuration.
  }
  \label{tab:deployment-cost}
  \small
  \resizebox{\columnwidth}{!}%
  {%
    \begin{tabular}{l r r r |l r r r}
      \toprule
      \multirow{2}[2]{*}{\textbf{Program}} & \textbf{Time}                        & \textbf{\#Tok.} & \textbf{Cost}
                                           & \multirow{2}[2]{*}{\textbf{Program}} & \textbf{Time}   & \textbf{\#Tok.} & \textbf{Cost} \\
                                           & (min)                                & (M)             & (\$)
                                           &                                      & (min)           & (M)             & (\$) \\
      \midrule
      bash                                 & 17.5                                 & 7.72            & 11.39           & bc       & 20.6 & 9.41 & 14.52 \\
      bison                                & 17.8                                 & 7.83            & 12.46           & cjpeg    & 25.5 & 4.34 & 8.22 \\
      cjson                                & 14.0                                 & 6.06            & 10.96           & combine  & 17.9 & 7.77 & 14.23 \\
      curl                                 & 18.5                                 & 8.16            & 11.46           & diff     & 25.4 & 7.41 & 13.94 \\
      expat                                & 20.3                                 & 6.64            & 11.11           & find     & 26.8 & 7.02 & 11.34 \\
      flex                                 & 22.6                                 & 6.33            & 11.89           & flvmeta  & 24.0 & 5.71 & 10.10 \\
      gawk                                 & 23.8                                 & 7.74            & 11.57           & grep     & 18.5 & 6.98 & 11.29 \\
      gzip                                 & 21.6                                 & 5.09            & 10.13           & jasper   & 25.9 & 7.32 & 14.62 \\
      jq                                   & 21.6                                 & 5.23            & 13.94           & tiffinfo & 20.7 & 6.38 & 12.29 \\
      xmllint                              & 17.8                                 & 4.07            & 8.98            & lua      & 18.8 & 3.95 & 8.40 \\
      make                                 & 19.3                                 & 5.19            & 11.18           & nasm     & 24.7 & 6.41 & 10.22 \\
      objcopy                              & 22.0                                 & 5.12            & 11.81           & openssl  & 19.4 & 8.15 & 11.66 \\
      patch                                & 19.5                                 & 5.71            & 11.95           & readelf  & 21.8 & 7.46 & 12.94 \\
      sed                                  & 24.4                                 & 4.61            & 7.84            & sqlite   & 24.8 & 5.64 & 10.73 \\
      strip-new                            & 23.5                                 & 5.26            & 10.75           & tic      & 21.4 & 8.61 & 13.92 \\
      transicc                             & 22.6                                 & 5.92            & 12.15           & vim      & 23.2 & 5.98 & 12.51 \\
      \midrule
      \textbf{Average}                     & 21.44                                & 6.41            & 11.58           & -        & -    & -    & - \\
      \bottomrule
    \end{tabular}%
  }
\end{table}

\symbolon incurs a recurring per-project cost before symbolic execution, \ie the agent retrieves learned skills, instantiates local rewrites, and validates candidate transformations.
\Cref{tab:deployment-cost} reports this deployment cost for applying the distilled skill library, excluding the one-time offline rule-learning phase.
Across the 32 programs, deployment takes 21.44 minutes, consumes 6.41M tokens, and costs \$11.58 on average.
We charge this time to \symbolon under the same 12-hour end-to-end budget used by the baselines.
On average, deployment consumes only 3.0\% of the budget, leaving the remaining time for symbolic exploration.
The worst-case time is 26.8 minutes on \texttt{find}, and the highest cost is \$14.62 on \texttt{jasper}.
As the transformed source is produced before symbolic execution and is independent of the engine configuration, it can be reused across repeated campaigns, different budgets, and, in our evaluation, all 16 search strategies.
Therefore, the recurring deployment overhead remains reasonably small relative to the symbolic execution campaigns it accelerates, while the more expensive open-ended transformation rule discovery is amortized through offline one-time skill learning.

\subsection{Sensitivity Analysis}
\label{sec:evaluation/sensitivity}

We further study how \symbolon behaves under different implementation choices and deployment settings.
Specifically, we vary the agent and LLM backend, ablate components of the learned skill library, and test whether the same high-level workflow transfers to other symbolic execution engines and programming languages.

\begin{table}[!t]
  \centering
  \caption{Sensitivity analysis of \symbolon to different settings.
  \emph{Agent and Models}: different agent setting for the online inference in \symbolon. \emph{Skill Library}: different settings for integrating the skill library to the agent. \textbf{LCov} is averaged line coverage across 32 benchmark programs with \texttt{nurs:covnew} search strategy; \(\Delta\) is the relative change compared to the full \symbolon\ (top row).}
  \label{tab:ablation-study}
  \small
  {%
    \begin{tabularx}{\columnwidth}{@{}>{\raggedright\arraybackslash\hspace*{1.2em}}X l l@{}}
      \toprule
      \textbf{Configuration}                               & \textbf{LCov} & \textbf{\(\Delta\) (\%)} \\
      \midrule
      \multicolumn{3}{@{}l}{\emph{\symbolon\ (default)}} \\
      \claudelogo\,Claude Code + Claude Opus 4.6           & 5,336         & --- \\
      \midrule
      \multicolumn{3}{@{}l}{\emph{Agent and Models}} \\
      \claudelogo\,Claude Code + Claude Sonnet 4.6         & 4,427         & -17.0 \\
      \codexlogo\,Codex + GPT 5.5 \emph{high effort}       & 6,590         & +23.5 \\
      \codexlogo\,Codex + GPT 5.3-Codex \emph{high effort} & 4,892         & -8.3 \\
      \midrule
      \multicolumn{3}{@{}l}{\emph{Skill Library}} \\
      w/o Pattern Augmentation                             & 2,951         & -44.7 \\
      w/o Transformation                                   & 1,468         & -72.5 \\
      w/o Barrier Description                              & 1,163         & -78.2 \\
      \bottomrule
    \end{tabularx}%
  }
\end{table}

\paragraph{Agent and LLM backends}
We first vary the agent and model used during repo-level transformation while keeping the learned skill library fixed.
As shown in \Cref{tab:ablation-study}, \symbolon remains effective across different backends, while their final coverage depends on model and agent quality.
Replacing the default Claude Code + Claude Opus~4.6 with Claude Sonnet~4.6 reduces average line coverage by 17.0\%, suggesting that stronger reasoning and code capability help adapt learned transformations to target programs.
Using Codex with GPT~5.5 in high-effort mode improves coverage by 23.5\% over the default, while GPT~5.3-Codex performs 8.3\% below the default but remains competitive.
These results show that \symbolon is not tied to a single LLM or agent runtime, while stronger backends can further improve repo-level transformation.

\paragraph{Skill library}
We next ablate the skill library to understand which learned information is most useful during repo-level transformation.
Removing pattern augmentation reduces average line coverage by 44.7\%, showing that additional trigger patterns help the agent recognize transformation opportunities across diverse target contexts.
Removing concrete before-and-after transformation examples causes a larger drop of 72.5\%, indicating that examples are critical for turning a barrier diagnosis into an actionable rewrite.
Removing the barrier description reduces coverage by 78.2\%, the largest degradation among the ablations.
This suggests that effective skill transfer requires both \emph{what} to transform and \emph{why} the code impedes symbolic execution.
Syntactic patterns or examples alone are insufficient for context-sensitive transformation.

\paragraph{Generalization to other languages}
\label{sec:evaluation/other-languages}

\symbolon operates at the source level and does not depend on specific intermediate representations, solver APIs, or search policies.
To test this generalizability, we integrate \symbolon with three additional symbolic execution engines, \ie \texttt{Owi}\cite{andres2024owi} for Rust, \texttt{CrossHair}\cite{schanely2022crosshair,bruni2011peer} for Python, and \texttt{ExpoSE}\cite{loring2017expose} for JavaScript.
As these engines expose different execution interfaces, we keep the experiment lightweight.
Specifically, \symbolon starts from the existing learned skill library and performs repo-level transformation, without engine-specific rule learning.
We evaluate two projects per language using each engine's default configuration, under the same 12-hour budget for both the baseline and the transformed program.
Each configuration is repeated three times, and coverage is measured by replaying generated inputs on the original program.
The evaluated projects and versions are listed in \Cref{app:programs}.

\Cref{tab:other-languages} summarizes the results.
\symbolon improves coverage for all three engines, with gains of 82\% for \texttt{Owi}, 178\% for \texttt{CrossHair}, and 35\% for \texttt{ExpoSE}.
The coverage increases by 56\% across the six projects.
These results are not intended as a full benchmark for non-C engines, but they provide evidence that the source-level transformation workflow can transfer beyond the default C/KLEE setting with minimal engine-specific engineering.

\begin{table}[!t]
  \centering
  \caption{Generalization of \symbolon\ across symbolic-execution
    engines and source languages. We report aggregate line coverage over each engine's benchmark projects.
    \(\Delta\): relative gain of \symbolon\ over running the engine alone.
  }
  \label{tab:other-languages}
  \small
  \setlength{\tabcolsep}{3pt}
  {%
    \begin{tabular*}{\columnwidth}{@{\extracolsep{\fill}}l l c rrr@{}}
      \toprule
      \multirow{2}{*}{\textbf{Engine}}     &
      \multirow{2}{*}{\textbf{Language}}   &
      \multirow{2}{*}{\textbf{\#Projects}} &
      \multicolumn{3}{c}{\textbf{Line Coverage}} \\
      \cmidrule(lr){4-6}
      &            &   & Vanilla & \symbolon & \(\Delta\)\,(\%) \\
      \midrule
      \texttt{Owi}       & \langicon{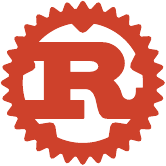}~Rust       & 2 & 171   & 312   & +82  \\
      \texttt{CrossHair} & \langicon{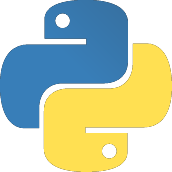}~Python          & 2 & 141   & 393   & +178 \\
      \texttt{ExpoSE}    & \langicon{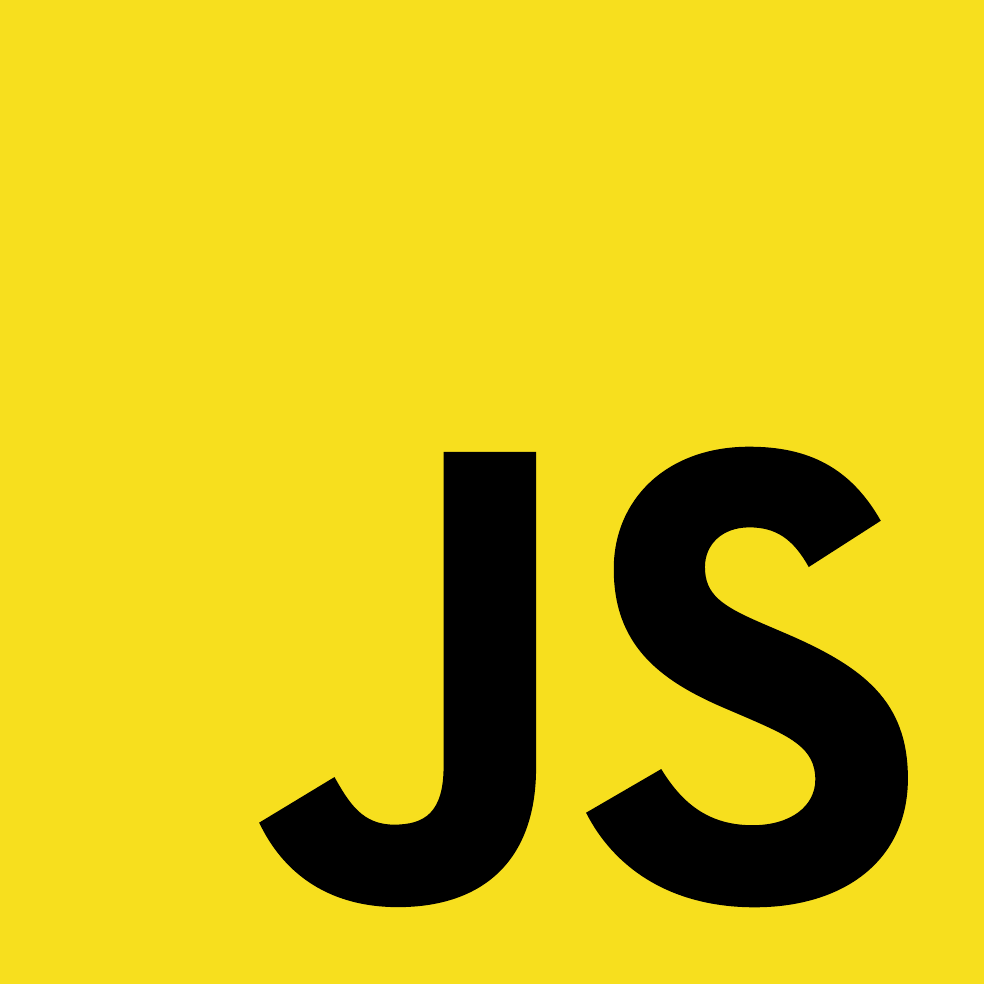}~JavaScript          & 2 & 1,079 & 1,464 & +35  \\
      \midrule
      \textbf{Total}                       &            & 6 & 1,391   & 2,169     & +56 \\
      \bottomrule
    \end{tabular*}%
  }
\end{table}

\section{Discussion}
\label{sec:discussion}

\paragraph{Cost of agentic transformation}
\symbolon amortizes open-ended rule discovery by learning transformations offline on small programs, but applying learned skills to a new repository still incurs a repo-level transformation cost.
The agent must locate candidate contexts, instantiate rules, validate transformed code, and run symbolic execution and replay a few times before the main campaign.
In our evaluation, we charge this repo-level transformation and validation time to \symbolon by reducing its remaining symbolic-execution budget, so \symbolon and the baselines are compared under the same end-to-end budget (\Cref{sec:evaluation/setup}).
However, this setting still does not rule out the possibility that a promising initial transformation fails to translate into improved performance over a longer campaign, in which case the transformation overhead can outweigh the eventual coverage gain.

\paragraph{Limited transformations applied}
We observe that in practice, the applied transformations per project are rare.
Across our benchmark programs, the agent uses only a few transformation types and applies each type a handful of times per program (\Cref{tab:program-composition}).
One potential explanation can be the limited types of program in our evaluation.
Most evaluated programs are file or command-line parsers, whose symbolic-execution challenges arise from a small set of recurring patterns, such as input-parsing loops.
In contrast, on the \texttt{logic-bomb} benchmark designed for symbolic execution \cite{xu2020benchmarking}, \symbolon resolves a broader range of challenges using more diverse and more intricate transformations (\Cref{tab:logic-bomb-programs}).
Another factor might be the limited inference budget allocated to the agent, as we reserve most of the fixed end-to-end budget for symbolic execution.
That said, note that even under this constrained setting, the resulting transformations produce nontrivial coverage gains (\Cref{sec:evaluation/rq1}).
The restricted budget prevents the agent from discovering additional or more diverse transformations that could further improve exploration.
\Cref{app:iterative-agent-analysis} provides preliminary evidence that allocating a larger inference budget to the agent can result in further coverage gains.
Future work could develop adaptive allocation policies that decide whether the next unit of budget is better spent searching for another transformation or continuing symbolic execution.

\paragraph{Learning--deployment gap}
\symbolon learns transformation skills from small benchmark programs because reward evaluation is cheap in that setting.
While our evaluation shows that these skills transfer to larger real-world projects, a gap can remain between offline-learning programs and deployment targets.
Real projects often include domain-specific APIs, macros, and long-range dependencies that may not appear in small coding benchmarks.
As a result, some real-world barriers may not be covered by our learned skill library, and some learned rules may need to be substantially adapted before becoming useful.
Future work could reduce this gap by learning from broader or domain-specific corpora, maintaining project-specific skill libraries, or developing cheaper proxy rewards for transformation search.

\section{Related Work}
\label{sec:related-work}

\paragraph{Optimizations for symbolic execution}
A large body of work improves symbolic execution by optimizing the engine, solver, or exploration policy.
Existing approaches employ search strategies and learned state-selection policies to prioritize promising program states\cite{li2013sgs,ruaro2021syml,he2021learch,yi2024cbc,yao2025empc}, and leverage pruning, slicing, memoization, and under-constrained execution to reduce the program region to explore or reuse prior reasoning\cite{siddiqui2012scaling,trabish2018chopped,yang2012memoized,ramos2015under}.
Solver-aware techniques use concrete executions, constraint simplification, or specialized reasoning to reduce query cost\cite{coppa2017rethinking,chen2018learning,sun2024cgs}.
Other systems improve symbolic execution engines for specific deployment settings, such as hybrid fuzzing, binary analysis, firmware analysis, and long-running campaigns\cite{yun2018qsym,wang2017angr,david2016binsec,qi2024symfit,liu2024co3,busse2020moklee}, leveraging domain-specific optimization opportunities.
Recent ML/LLM-assisted symbolic execution further optimizes the analysis loop to generate tests, interpret or solve path constraints, complete structured inputs, summarize constraints, or guide vulnerability discovery\cite{xu2024symbolic,cha2020making,li2025large,wu2025generating,xia2026symgpt,yang2025hlpfuzz,tu2026cottontail,luo2026concollmic,shafiuzzaman2026guiding}.
\symbolon complements these approaches by changing the source representation before these approaches take place, while staying disentangled from all optimizations during symbolic execution.

\paragraph{Program transformation for symbolic execution}
A complementary direction reduces symbolic execution cost by enriching the program representation (via compilation and transformation) exposed to the symbolic execution engine\cite{poeplau2020symcc,wei2023compiling}.
As program analyses are known to be sensitive to code transformations\cite{van2020tailoring,namjoshi2018impact,peng2018tfuzz,guler2019antifuzz,jung2019fuzzification,zhu2025locus,luo2026code}, existing approaches have also exploited this to improve symbolic execution\cite{cadar2015targeted,perry2017accelerating,dong2015studying,zhang2024compiler,chen2018learning,converse2017non,saumya2026taming,zhu2024loopscc,barr2018indexing,bouras2026defusing}.
However, their transformation spaces are usually fixed by compiler passes or narrow expert-written templates tied to specific program constructs.
\symbolon builds on a similar insight but treats transformation discovery as a learning problem.
It automatically searches for diverse transformation rules from symbolic execution feedback, stores them as persistent agent skills, and instantiates them across diverse contexts in the project to optimize symbolic execution.

\paragraph{LLM-guided code evolution}
Recent work has shown that LLMs can serve not only as code generators but also as proposal mechanisms in efficient search over programs, algorithms, and agentic systems\cite{zhang2025darwin,xia2025live, novikov2025alphaevolve,romera2024mathematical,sen2026kiss}.
For example, AlphaEvolve and ADRS use evolutionary coding-agent loops to find state-of-the-art algorithms and system architectures that human experts have missed for decades\cite{novikov2025alphaevolve,cheng2025barbarians}.
These approaches suggest a new form of learning in which knowledge is accumulated through code and agent memory/skills, rather than solely through updates to model weights.
\symbolon follows a similar spirit, but differs in both the optimization target and the deployment model.
Rather than evolving the program to improve its own performance, \symbolon evolves program representations that improve how symbolic execution reasons about them.
Moreover, \symbolon does not run expensive evolution on every large repository.
It performs cheap transformation discovery offline on small programs, and transfers the learned knowledge via persistent agent skills\cite{xu2026agent,anthropic2025agentskills} to repo-level transformations.
Therefore, \symbolon suggests a broader future direction for evolutionary search when reward computation is expensive, making it promising to extend program representation learning to other code reasoning tasks, \eg fuzzing, formal verification, and agentic software analysis.

\section{Conclusion}
\label{sec:conclusion}

This paper introduces \symbolon, a framework that learns source-level program transformations to improve symbolic execution.
Rather than relying on fixed compiler optimizations or hand-written rewrite rules, \symbolon automatically searches for program representations amenable to symbolic execution, validates them by replaying the tests on the original program, and learns transformation rules as persistent agent skills for agentic, repo-level transformation on popular real-world software projects.
By measuring progress on the original program, \symbolon keeps the optimization target aligned with the goal of symbolic execution, while allowing a broad transformation space.
Our evaluation across 32 real-world C programs and 16 search strategies shows consistent improvements in coverage, with substantially reduced overhead in symbolic execution.
\symbolon also finds new security violations in these programs, as well as Linux kernel bugs when used to augment state-of-the-art kernel bug finders based on symbolic execution.

\section*{Acknowledgments}
This research is in part based upon work supported by Coefficient Giving and by credits supported through Amazon Nova AI Challenge: Trusted Software Agents.
Results presented in this paper were obtained using the Chameleon testbed supported by the National Science Foundation.
Any opinions, findings, and conclusions or recommendations expressed in this material are those of the author(s) and do not necessarily reflect the views of the sponsors.
Language model assistants were used to help polish the paper.

\bibliographystyle{IEEEtran}
\bibliography{ref}

\appendices
\crefalias{section}{appendix}

\newcommand{\appsnippet}[2]{%
  \ifusepdfcode
  \par\nobreak\vspace{4pt}\noindent\codepdf{#2}\par\vspace{4pt}
  \else
  \begin{codebox}[snippet, language=cpp, /tcb/.cd, before skip=4pt, after skip=4pt]{#1}
  \end{codebox}%
\fi}

\newlength{\transgap}\setlength{\transgap}{15pt}%
\newif\iftransfirst
\newcommand{\transformation}[8]{%
  \par
  \iftransfirst\transfirstfalse\medskip\else\vspace{\transgap}\fi
  \noindent
  \begingroup\renewcommand{\arraystretch}{1.3}%
  \begin{tabular}{|p{\dimexpr\linewidth-2\tabcolsep-2\arrayrulewidth\relax}|}
    \hline
    \textbf{Name:}~\texttt{#1} \tabularnewline
    \hline
    \textbf{Trigger:}~#2 \tabularnewline
    \hline
    \textbf{Transformation:}~#3 \tabularnewline
    \hline
    \textbf{Rationale:}~#4 \tabularnewline
    \hline
    \textbf{Before}\par\nobreak\appsnippet{#5}{#6} \tabularnewline
    \hline
    \textbf{After}\par\nobreak\appsnippet{#7}{#8} \tabularnewline
    \hline
  \end{tabular}\par
  \endgroup
}

\section{KLEE Command-Line Configuration}
\label{app:klee-command}

The evaluation uses the following KLEE options:

\begin{verbatim}
--solver-backend=z3 --max-memory=32768 --watchdog \
--only-output-states-covering-new --posix-runtime \
--libc=uclibc --dump-states-on-halt=false \
--write-kqueries --write-smt2s --write-cov
\end{verbatim}

\section{Per-program Symbolic Environments}
\label{app:program-symenv}

\Cref{tab:program-symenv} lists the exact per-program argument sequence appended after the program bitcode in each KLEE invocation.
Literal tokens are passed to the target as concrete command-line arguments, whereas the \texttt{--sym-*} options are interpreted by KLEE's POSIX runtime.

\begin{table*}[!t]
  \centering
  \caption{Exact symbolic-environment and concrete program arguments used for
    the 32 evaluated programs. Each sequence is appended after the program
    bitcode; order is preserved from the corresponding \texttt{symEnv}
  definition.}
  \label{tab:program-symenv}
  \small
  \setlength{\tabcolsep}{3pt}
  \renewcommand{\arraystretch}{1.08}
  \begin{tabularx}{\textwidth}{@{}l >{\raggedright\arraybackslash\ttfamily}X | l >{\raggedright\arraybackslash\ttfamily}X@{}}
    \toprule
    \textbf{Program} & \normalfont\textbf{Symbolic Environment} &
    \textbf{Program} & \normalfont\textbf{Symbolic Environment} \\
    \midrule
    \texttt{bash} & --sym-args 0 3 10 --sym-stdin 100 &
    \texttt{bc} & --sym-stdin 40 \\
    \texttt{bison} & --sym-args 0 2 2 A --sym-files 1 100 &
    \texttt{cjpeg} & --sym-args 0 2 4 A --sym-files 1 100 \\
    \texttt{cjson} & A --sym-files 1 100 yes &
    \texttt{combine} & --sym-args 0 2 4 A B --sym-files 2 50 \\
    \texttt{curl} & --sym-args 0 4 6 --sym-args 0 1 20 &
    \texttt{diff} & --sym-args 0 2 2 A B --sym-files 2 50 \\
    \texttt{expat} & --sym-stdin 100 &
    \texttt{find} & --sym-args 0 3 10 --sym-files 1 40 --sym-stdin 40 \\
    \texttt{flex} & --sym-args 0 5 12 A --sym-files 1 300 &
    \texttt{flvmeta} & --sym-arg 2 --sym-args 0 2 6 A --sym-files 1 300 --save-all-writes \\
    \texttt{gawk} & -f A B --sym-files 2 50 &
    \texttt{grep} & --sym-args 0 2 2 --sym-arg 10 A --sym-files 1 50 \\
    \texttt{gzip} & --sym-args 0 2 2 A --sym-files 1 50 &
    \texttt{jasper} & --input A --output B --input-format --sym-arg 3 --output-format --sym-arg 3 --sym-args 0 3 15 --sym-files 2 300 --save-all-writes \\
    \texttt{jq} & --sym-args 1 1 10 --sym-stdin 50 &
    \texttt{xmllint} & A --sym-files 1 40 \\
    \texttt{lua} & --sym-stdin 64 &
    \texttt{make} & -n -f A --sym-files 1 40 \\
    \texttt{nasm} & --sym-args 0 2 2 A --sym-files 1 100 &
    \texttt{objcopy} & --sym-args 0 3 20 A --sym-files 1 100 \\
    \texttt{openssl} & asn1parse -inform DER --sym-stdin 64 &
    \texttt{patch} & --sym-args 0 2 2 A B --sym-files 2 50 \\
    \texttt{readelf} & -a A --sym-files 1 100 &
    \texttt{sed} & --sym-arg 10 --sym-args 0 2 2 --sym-files 1 8 --sym-stdin 8 \\
    \texttt{sqlite} & --sym-stdin 20 --sym-stdout &
    \texttt{strip-new} & --sym-args 0 2 8 A --sym-files 1 100 \\
    \texttt{tic} & --sym-args 0 2 8 A --sym-files 1 100 &
    \texttt{tiffinfo} & --sym-args 0 3 8 A --sym-files 1 300 \\
    \texttt{transicc} & --sym-args 0 2 4 A B --sym-files 2 200 --save-all-writes &
    \texttt{vim} & -N -E -s --sym-stdin 120 \\
    \bottomrule
  \end{tabularx}
\end{table*}

\section{Projects for Evaluating Generalization}
\label{app:programs}

\Cref{tab:gen-projects} details the projects behind the per-language results in
\Cref{tab:other-languages}, with their versions and sources.

\begin{table}[H]
  \centering
  \caption{Projects used to evaluate cross-language
  generalization (\Cref{tab:other-languages}).}
  \label{tab:gen-projects}
  \small
  \setlength{\tabcolsep}{2pt}
  \begin{tabularx}{\columnwidth}{@{}l l >{\raggedright\arraybackslash}X@{}}
    \toprule
    \textbf{Language} & \textbf{Project} & \textbf{Source} \\
    \midrule
    \multirow{2}{*}{\langicon{rust-logo-orange.pdf}~Rust}
    & \texttt{semver} (1.0.28)    & {\href{https://crates.io/crates/semver}{\nolinkurl{crates.io/crates/semver}}} \\
    & \texttt{humantime} (2.1.0)  & {\href{https://crates.io/crates/humantime}{\nolinkurl{crates.io/crates/humantime}}} \\
    \midrule
    \multirow{2}{*}{\langicon{python-logo.pdf}~Python}
    & \texttt{tomllib} (stdlib) & {\href{https://docs.python.org/3/library/tomllib.html}{\nolinkurl{docs.python.org/3/library/tomllib}}} \\
    & \texttt{json} (stdlib)    & {\href{https://docs.python.org/3/library/json.html}{\nolinkurl{docs.python.org/3/library/json}}} \\
    \midrule
    \multirow{2}{*}{\langicon{js-logo.pdf}~JavaScript}
    & \texttt{mathjs} (15.2.0) & {\href{https://www.npmjs.com/package/mathjs}{\nolinkurl{www.npmjs.com/package/mathjs}}} \\
    & \texttt{css} (3.0.0)    & {\href{https://www.npmjs.com/package/css}{\nolinkurl{www.npmjs.com/package/css}}} \\
    \bottomrule
  \end{tabularx}
\end{table}

\section{Transformation Composition per Program}
\label{app:program-composition}

For each of the 32 benchmark programs, \Cref{tab:program-composition} reports the number of transformation instances in the final \symbolon-transformed program, together with a per-transformation breakdown.
We treat each diff hunk as the minimum counting unit.
The reported frequencies may therefore underestimate the actual number of transformations, as a single hunk may contain multiple applications of the same transformation or several distinct transformations.

\begin{table*}[t]
  \centering
  \caption{Per-program transformation composition: the number of distinct
    source-level transformation instances (\#) applied in each program's final
    \symbolon-transformed version and their per-transformation breakdown classified from each patch's recorded rationale.
  Program names and ordering follow \Cref{tab:program-stats}.}
  \label{tab:program-composition}
  \small
  \setlength{\tabcolsep}{4pt}
  \begin{tabularx}{\textwidth}{@{}l c >{\raggedright\arraybackslash}X | l c >{\raggedright\arraybackslash}X@{}}
    \toprule
    \textbf{Program} & \textbf{\#} & \textbf{Transformation Breakdown} &
    \textbf{Program} & \textbf{\#} & \textbf{Transformation Breakdown} \\
    \midrule
    \texttt{bash} & 3 & \texttt{loop-bound}\,(2), \texttt{noexec-guard} & \texttt{bc} & 3 & \texttt{lookup-to-dataflow}\,(2), \texttt{solver-hotspot-stub} \\
    \texttt{bison} & 3 & \texttt{loop-bound}\,(2), \texttt{reuse-input-var} & \texttt{cjpeg} & 2 & \texttt{format-dispatch-pin}, \texttt{valid-image-pin} \\
    \texttt{cjson} & 1 & \texttt{loop-bound} & \texttt{combine} & 2 & \texttt{loop-bound}, \texttt{state-killer-stub} \\
    \texttt{curl} & 2 & \texttt{cmd-prefix-bound}, \texttt{recur-to-iter} & \texttt{diff} & 3 & \texttt{loop-bound}, \texttt{input-prefix-pin}, \texttt{stat-pin} \\
    \texttt{expat} & 2 & \texttt{loop-bound}, \texttt{menu-pin} & \texttt{find} & 2 & \texttt{array-to-scalar-vars}, \texttt{string-compare} \\
    \texttt{flex} & 4 & \texttt{loop-bound}\,(2), \texttt{state-killer-stub}, \texttt{out-bypass} & \texttt{flvmeta} & 3 & \texttt{loop-bound}, \texttt{menu-pin}, \texttt{concretize-hash} \\
    \texttt{gawk} & 2 & \texttt{loop-bound}\,(2) & \texttt{grep} & 3 & \texttt{loop-bound}\,(2), \texttt{switch-lowering} \\
    \texttt{gzip} & 2 & \texttt{concretize-hash}, \texttt{state-killer-stub} & \texttt{jasper} & 2 & \texttt{menu-pin}, \texttt{recur-to-iter } \\
    \texttt{jq} & 2 & \texttt{loop-bound}, \texttt{switch-lowering} & \texttt{xmllint} & 2 & \texttt{switch-lowering}, \texttt{symbolic-bound-mem} \\
    \texttt{lua} & 3 & \texttt{loop-bound}, \texttt{tty-reroute}, \texttt{longjmp-model} & \texttt{make} & 3 & \texttt{loop-bound}, \texttt{state-killer-stub}\,(2) \\
    \texttt{nasm} & 3 & \texttt{loop-bound}, \texttt{symbolic-bound-mem}, \texttt{menu-pin} & \texttt{objcopy} & 2 & \texttt{input-prefix-pin}, \texttt{valid-image-pin} \\
    \texttt{openssl} & 3 & \texttt{loop-bound}\,(2), \texttt{menu-pin} & \texttt{patch} & 2 & \texttt{loop-bound}, \texttt{input-prefix-pin} \\
    \texttt{readelf} & 1 & \texttt{input-prefix-pin} & \texttt{sed} & 2 & \texttt{loop-bound}, \texttt{input-prefix-pin} \\
    \texttt{sqlite} & 2 & \texttt{string-compare}, \texttt{switch-lowering} & \texttt{strip-new} & 2 & \texttt{input-prefix-pin}, \texttt{avoid-bound-sentinels} \\
    \texttt{tic} & 2 & \texttt{concretize-hash}, \texttt{input-prefix-pin} & \texttt{tiffinfo} & 3 & \texttt{loop-bound}, \texttt{symbolic-bound-mem}, \texttt{io-model} \\
    \texttt{transicc} & 3 & \texttt{loop-bound}, \texttt{float-to-int}, \texttt{loop-removing} & \texttt{vim} & 3 & \texttt{io-model}\,(2), \texttt{avoid-loop-based-search} \\
    \bottomrule
  \end{tabularx}
\end{table*}

\section{Coverage Gains with Longer Inference Time}
\label{app:iterative-agent-analysis}

We conduct additional experiments on \texttt{bc} and \texttt{flvmeta} to study how the \symbolon inference-time budget affects coverage improvement.
For each program, we run \symbolon with inference budgets of 10, 20, 30, and 40 minutes while keeping the total end-to-end evaluation budget fixed at 12 hours across all settings.
We then evaluate the resulting transformed program using the replay-based coverage procedure described in \Cref{sec:evaluation/setup}.

As shown in \Cref{tab:agent-effort-coverage}, coverage improves at every tested budget for both programs, suggesting that additional inference time allows \symbolon to discover and apply further coverage-enhancing transformations.

\begin{table}[H]
  \centering
  \caption{Line coverage of \texttt{bc} and \texttt{flvmeta} under increasing
    \symbolon inference time budgets. The vanilla column reports the coverage without applying \symbolon.}
  \label{tab:agent-effort-coverage}
  \small
  \setlength{\tabcolsep}{4pt}
  \begin{tabularx}{\columnwidth}{@{}l *{5}{>{\centering\arraybackslash}X}@{}}
    \toprule
    \multirow{2}{*}{\textbf{Program}} & \multicolumn{5}{c}{\textbf{Covered LoC by inference budget}} \\
    \cmidrule(l){2-6}
    & \textbf{Vanilla} & \textbf{10min} & \textbf{20min} & \textbf{30min} & \textbf{40min} \\
    \midrule
    \texttt{bc}      & 660   & 773   & 953   & 1,384 & 1,615 \\
    \texttt{flvmeta} & 1,426 & 1,970 & 2,080 & 3,354 & 3,508 \\
    \bottomrule
  \end{tabularx}
\end{table}

\section{Evaluations on Logic-Bomb Benchmarks}
\label{app:logic-bomb-transformations}

We further evaluate \symbolon on the logic-bomb symbolic-execution benchmark\cite{xu2020benchmarking}.
This benchmark contains 53 small C/C++ programs for KLEE covering 12 different categories.
For each program, we follow the same experiment setting as in \Cref{sec:evaluation/setup} with a 60-second time budget.
\symbolon applies diverse transformation skills and achieves 100\% coverage on 50 of the 53 programs, covering 38 more programs than vanilla symbolic execution.
The remaining three programs cannot be fully covered because the target flag resides in dead code in \texttt{stack\_bo\_l2}, depends on an unmodeled process identifier in \texttt{pid\_csv}, or is absent from \texttt{file\_posix\_cp\_l1}.
\Cref{tab:logic-bomb-programs} lists the 38 successfully triggered logic bombs, together with the transformation skills \symbolon applied.

\begin{table*}[!t]
  \centering
  \caption{The successfully triggered \texttt{logic-bomb} benchmark programs\cite{xu2020benchmarking}
    evaluated with \symbolon, grouped by category. \textbf{\#}: number of triggered
  programs in the category.}
  \label{tab:logic-bomb-programs}
  \footnotesize
  \setlength{\tabcolsep}{3pt}
  \renewcommand{\arraystretch}{1.05}
  \renewcommand{\tabularxcolumn}[1]{>{\raggedright\arraybackslash}m{#1}}
  \begin{tabularx}{\textwidth}{@{}l c X X@{}}
    \toprule
    \textbf{Category} & \textbf{\#} & \textbf{Programs} & \textbf{Used Transformations} \\
    \midrule
    \texttt{buffer\_overflow} & 3 &
    \texttt{heap\_bo\_l1}, \texttt{stack\_bo\_l1},
    \texttt{stacknocrash\_bo\_l1} &
    \texttt{explicit-adjacent-memory-model}, \texttt{model-overflow-into-field}, \texttt{heap-to-global-aliasing-model} \\
    \midrule
    \texttt{contextual\_symbolic\_value} & 2 &
    \texttt{ping\_csv}, \texttt{syscall\_csv} &
    \texttt{unmodeled-syscall-as-nondeterministic-input} \\
    \midrule
    \texttt{covert\_propogation} & 6 &
    \texttt{echo\_cp\_l1}, \texttt{echofile\_cp\_l1},
    \texttt{file\_cp\_l1}, \texttt{file\_posix\_cp\_l1}, \texttt{socket\_cp\_l1},
    \texttt{stack\_cp\_l1} &
    \texttt{inline-asm-to-equivalent-c}, \texttt{in-memory-string-conversion}, \texttt{in-memory-io-roundtrip}, \texttt{external-process-to-in-memory-model}, \texttt{loopback-socket-to-in-memory-buffer} \\
    \midrule
    \texttt{crypto\_functions} & 2 &
    \texttt{aes\_cf}, \texttt{sha\_cf} &
    \texttt{finite-domain-preimage-enumeration}, \texttt{algebraic-key-schedule-inversion} \\
    \midrule
    \texttt{external\_functions} & 7 &
    \texttt{atof\_ef\_l2}, \texttt{ln\_ef\_l2}, \texttt{pow\_ef\_l2}, \texttt{printfloat\_ef\_l1}, \texttt{printint\_int\_l1},
    \texttt{rand\_ef\_l2}, \texttt{sin\_ef\_l2} &
    \texttt{symbolic-to-concrete-case-split}, \texttt{external-prng-concretization-loop}, \texttt{sink-external-call-after-branch}, \texttt{concrete-case-split-on-input-byte}, \texttt{float-to-integer-arithmetic}, \texttt{concretize-finite-domain-float-call}, \texttt{symbolic-integer-decimal-compare} \\
    \midrule
    \texttt{floating\_point} & 5 &
    \texttt{float1\_fp\_l1}, \texttt{float2\_fp\_l1}, \texttt{float3\_fp\_l2},
    \texttt{float4\_fp\_l2}, \texttt{float5\_fp\_l2} &
    \texttt{finite-domain-float-concretization}, \texttt{float-to-integer-branch-decision}, \texttt{inline-integer-mantissa-annotation}, \texttt{float-free-decision-via-threshold-tables}, \texttt{concrete-indexed-accepting-interval-scan} \\
    \midrule
    \texttt{loop} & 5 &
    \texttt{5n+1\_lo\_l1}, \texttt{7n+1\_lo\_l1}, \texttt{collaz\_lo\_l1},
    \texttt{collaz\_lo\_l2}, \texttt{paraloop\_lo\_l2} &
    \texttt{input-domain-concretization}, \texttt{input-domain-enumeration}, \texttt{byte-concretization-fork}, \texttt{small-domain-input}, \texttt{concurrency-to-nondeterministic-schedule} \\
    \midrule
    \texttt{parallel\_program} & 5 &
    \texttt{2thread\_pp\_l1}, \texttt{2thread\_pp\_l2}, \texttt{forkpipe\_pp\_l1},
    \texttt{forkshm\_pp\_l1}, \texttt{mthread\_pp\_l2} &
    \texttt{thread-schedule-sequentialization}, \texttt{thread-race-analytic-outcome-set}, \texttt{model-opaque-ipc-roundtrip-as-direct-copy}, \texttt{thread-to-sequential-schedule} \\
    \midrule
    \texttt{symbolic\_jump} & 2 &
    \texttt{arrayjmp\_sj\_l2}, \texttt{jmp\_sj\_l1} &
    \texttt{longjmp-model}, \texttt{state-killer-stub} \\
    \midrule
    \texttt{symbolic\_memory} & 1 &
    \texttt{heapoutofbound\_sm\_l2} &
    \texttt{symbolic-index-to-explicit-select} \\
    \bottomrule
  \end{tabularx}
\end{table*}

\section{Sample Transformations}
\label{app:transformation_table}

\symbolon's offline learning phase
(\Cref{sec:methodology/transform-rule-learning}) produces \(211\) transformation rules, each of which is packaged as an agent skill that records a natural-language trigger, the transformation it performs, the reason it helps symbolic execution, and a verified before/after C example.
We present a few representative rules below.
Interestingly, some of them have also been found useful in recent work\cite{saumya2026taming}, yet \symbolon discovered them through automated learning without relying on any human expert.

\raggedbottom
\transfirsttrue

\transformation{lower-symbolic-lookup-to-dataflow}{%
  A \emph{symbolic} value indexes a constant table whose entry encodes which class the value falls into, and a branch then tests that entry.%
}{%
  Recast the lookup as data flow: a single branchless arithmetic predicate over
  the index (bitwise \texttt{\&}/\texttt{|}, never \texttt{||}/\texttt{\&\&})
  evaluated on one path.%
}{%
  Symbolic indexing is control flow in disguise; KLEE must encode the entire
  table through the SMT theory of arrays. \emph{Coverage gain:} \(+38\%\).%
}{assets/code/appendix/symbolic-array-index-before.c}{appendix-symbolic-array-index-before.pdf}%
{assets/code/appendix/symbolic-array-index-after.c}{appendix-symbolic-array-index-after.pdf}

\transformation{avoid-nonlinear-symbolic-constraints}{%
  A branch condition is a nonlinear function of symbolic inputs, such as a
  product or square of symbolic variables.%
}{%
  Offload the nonlinear term to a concrete computation so the residual
  comparison is effectively linear in the symbolic inputs.%
}{%
  KLEE's SMT backend models such nonlinear arithmetic imprecisely or times out
  on it. \emph{Coverage gain:} \(+54\%\).%
}{assets/code/appendix/nonlinear-before.c}{appendix-nonlinear-before.pdf}%
{assets/code/appendix/nonlinear-after.c}{appendix-nonlinear-after.pdf}

\transformation{precompute-recursion-to-table}{%
  A value is computed by multi-way (branching) recursion that is re-evaluated
  for each input.%
}{%
  Precompute the recurrence into a table once, reducing each query to a
  concrete, constant-index lookup.%
}{%
  The per-input recursion multiplies path constraints and explodes the call
  tree. \emph{Coverage gain:} \(+10\%\).%
}{assets/code/appendix/recursion-before.c}{appendix-recursion-before.pdf}%
{assets/code/appendix/recursion-after.c}{appendix-recursion-after.pdf}

\transformation{avoid-large-fixed-arrays}{%
  A large fixed-size array is pre-allocated and populated as scratch storage,
  then accessed at input-dependent indices.%
}{%
  Replace the array with on-the-fly computation of each value when it is
  needed, removing the bulk storage.%
}{%
  The large array creates complex symbolic memory constraints and many
  symbolic-indexed accesses. \emph{Coverage gain:} \(+61\%\).%
}{assets/code/appendix/large-arrays-before.c}{appendix-large-arrays-before.pdf}%
{assets/code/appendix/large-arrays-after.c}{appendix-large-arrays-after.pdf}

\transformation{bitmask-instead-of-array-flags}{%
  Set membership or per-element boolean flags are stored in an auxiliary array
  and queried by indexed lookups, often inside a loop.%
}{%
  Encode the set as bits of an integer and test membership with a single
  branchless bitwise \texttt{AND}.%
}{%
  KLEE forks on every symbolic-indexed load and data-dependent branch. \emph{Coverage gain:} \(+87\%\).%
}{assets/code/appendix/bitmask-before.c}{appendix-bitmask-before.pdf}%
{assets/code/appendix/bitmask-after.c}{appendix-bitmask-after.pdf}

\end{document}